\documentclass[%
 aip,
 amsmath,amssymb,
 reprint,%
floatfix]{revtex4-1}

\usepackage{graphicx}
\usepackage{dcolumn}
\usepackage{xcolor}
\usepackage{bm}

\usepackage[utf8]{inputenc}
\usepackage[T1]{fontenc}
\usepackage{mathptmx}

\DeclareMathOperator\sgn{sgn}
\draft
\begin{document}

\title[Underscreening and hidden ion structures in large scale simulations of concentrated electrolytes]{Underscreening and hidden ion structures in large scale simulations of concentrated electrolytes}
\author{Emily Krucker-Velasquez}
\author{James W. Swan}%
 \email{jswan@mit.edu}
\affiliation{%
 Department of Chemical Engineering, Massachusetts Institute of Technology, Cambridge, Massachusetts 02139, USA
}%
\date{\today}
\begin{abstract}
 The electrostatic screening length predicted by Debye-H\"uckel theory decreases with increasing ionic strength, but recent experiments have found that the screening length can instead increase in concentrated electrolytes. This phenomenon, referred to as underscreening, is believed to result from ion-ion correlations and short-range forces such as excluded volume interactions among ions. We use Brownian Dynamics to simulate a version of the Restrictive Primitive Model for electrolytes over a wide range of ion concentrations, ionic strengths, and ion excluded volume radii for binary electrolytes. We measure the decay of the charge-charge correlation among ions in the bulk, and compare it against scaling trends found experimentally and determined in certain weak coupling theories of ion-ion correlation. Moreover, we find that additional large scale ion structures emerge at high concentrations. In this regime, the frequency of oscillations computed from the charge-charge correlation function is not dominated by electrostatic interactions but rather by excluded volume interactions and with oscillation periods on the order of the ion diameter. We also find that the nearest neighbor correlation of ions sharing the same charge transitions from negative at small concentrations to positive at high concentrations, representing the formation of small, like-charge ion clusters. We conclude that the increase in local charge density due to the formation of these clusters and the topological constraints of macroscopic charged surfaces can help explain the degree of underscreening observed experimentally.
\end{abstract}

\maketitle

\section{\label{sec:Intro} Introduction}
\indent The study of concentrated electrolytes has recently drawn considerable interest due to their central importance in various applications, ranging from colloidal self-assembly \cite{Russel1989, Israelachvili1991} and biological processes \cite{Eisenberg2013} to supercapacitors and batteries. \cite{Capacitators}  The delicate balance of long-range electrostatic interactions and steric repulsion poses a physically complex problem. The structures of ions in bulk and near interfaces dictate properties such as capacitance \cite{Capacitance} and the effective forces between colloids in solution. \cite{Colloidal_stability} It follows that a fundamental understanding of bulk structural properties and the decay of ion-ion correlations is vital for modern day applications.\\
\indent Ions immersed in a fluid emanate an electric field that attracts oppositely charged species, commonly referred to as counter-ions.\cite{Russel1989} This effect gives rise to screening: the charge density surrounding an ion is arranged such that the net electrostatic potential due to the ion and counter-ion distribution decays much faster than the bare Coulomb potential.\cite{hansen_mcdonald_2006} The process of structuring counter-ions and ions in solution is collective and many-bodied and determines the effective behavior of the electrolyte. In a dilute solution, electrostatic interactions are well described by the linearized Poisson-Boltzmann equation and Debye-H\"uckel (DH) theory, which dictates that interactions between charges decay exponentially with the distance, ${r}$, as ${\exp(-r/ \lambda_{ \mathrm{D} })}$. In this regime, the spatial decay of charge-charge correlations is given by the Debye screening length ${\lambda_{\mathrm{D}}= \sqrt{ \varepsilon_{\mathrm{f}} k_{\mathrm{B}}T  /(e^2 \sum_{a=1}^M n_\nu z_\nu^2 ) }}$, or its corresponding wave vector, $\kappa_{\mathrm{D}} = \lambda_{ \mathrm{D }}^{-1}$; where ${n_{\nu}}$ and ${z_{\nu}}$ are the bulk number density and valence of species ${\nu}$ respectively, ${e}$ is the fundamental charge, ${\epsilon_{\mathrm{f}}}$ is the solvent permittivity, and ${k_{\mathrm{B}}T}$ is the thermal energy.  $ \kappa_{\mathrm{D}} $ is a fundamental property of the electrolyte based only on its composition and the temperature and can be thought of as a proxy for the ionic strength of the solution.\\
\indent The hallmark of Debye-H\"uckel theory is its ideal solution assumption: ions behave as point charges. By definition, they occupy an infinitesimally small volume, and, consequently, charges surrounding a central ion can be represented by an averaged cloud of continuous charge density with a minimum distance of closest approach.\cite{Russel1989} In reality, ions are not points and, as their concentration increases, short-ranged repulsive (excluded volume) interactions compete with electrostatic forces when establishing the solution microstructure, making the DH approximation ill-equipped to describe the charge density profile in concentrated electrolytes. Additionally, the competition between packing and screening effects leads to an ordering of charge, which typically manifests as alternating sign of local charge density around a central ion due to the oppositely charged ions filling successive coordination shells.\\
\indent Efforts to understand and progress beyond the limitations of DH theory at high concentrations have grown in recent years.\cite{Israelachvili1991, Tadros2011, Zohar2006, Besteman2004} Smith et al. \cite{Smith2016} used a surface force balance (SFB) apparatus to measure the decay length in various electrolyte solutions and ionic liquids. Contrary to what was suggested by the classical, dilute limit theory, the measured decay length of the long-range component of the surface force, denoted $ 1 / \kappa $, was found to depend non-monotonically on the ion concentration.  For low ionic strengths, they found that $ \kappa \sim \kappa_{\mathrm{D}} $.  However, beyond the dilute limit, it was found that ${\kappa \sim 1/\kappa_{\mathrm{D}}^2}$ such that an increase in ionic strength increases the correlation length.  An additional length scale was needed to establish dimensional consistency in this scaling relationship. Perkin's group found that across a wide variety of salts, this length scale was given by the characteristic physical size of the ions.  This anomalously long screening length at high ionic strengths is usually referred to as ``underscreening,'' a phenomenon whose origin is believed to be electrostatic in nature. \cite{Lee2017,Gebbie2017}  We will hereafter use the term underscreening to describe all observations of correlation decay lengths that grow larger with increases in $ \kappa_D $. \\
\indent Before describing some analytical and computational results in the existing literature, it is worth discussing the magnitude of physically realizable Debye lengths.  Equivalently, we may ask, just how big can $ \kappa_{\mathrm{D}} $ get?  Consider a 1:1 electrolyte with ions occupying a volume fraction $ \phi_a $ associated with characteristic ion radius ${a}$.  Then we can rewrite the inverse Debye length as $ \kappa_{\mathrm{D}} a = \sqrt{ 3 \phi_a \lambda_\mathrm{B} / a } $, where $ \lambda_\mathrm{B} $ is the Bjerrum length, the length scale at which the thermal energy is the same magnitude as the electrostatic interactions between two elementary charges in the medium.  For water at room temperature, the Bjerrum length is about 0.7 nm, whereas when using the vacuum permittivity, it is almost 60 nm.  If the characteristic ion size $ a $ reflects the limits of packing for spheres, then a representative volume fraction of $ \phi_a = 0.63 $ can be chosen to reflect random close packing.  With this estimate, the biggest value of $ \kappa_{\mathrm{D}} a $ realizable at room temperature with a 1:1 electrolyte is about 10.  With a 2:2 electrolyte, this number increases by another factor of 2.  In the experiments of Smith \emph{et al}., \cite{Smith2016} across a wide range of electrolytes and ionic liquids, the Debye lengths are bounded by this same limiting scale. \\
\indent Mean-field, Poisson-Boltzmann type models have emerged to account for interactions omitted in DH approximation (such as ion-ion coupling and steric interactions). \cite{Onsager1933, Kjellander1992, BSK2011, Carvalho1994} In 1992, Kjellander proposed the reformulation of charged species in terms of quasiparticles or ``dressed'' ions.\cite{Kjellander1992} Moreira and Netz used statistical field theory to model ions near a charged surface in the limit of high counterion valence, surface charge and low temperatures.\cite{Moreira2000,Moreira2001} Bazant, Storey, and Kornyshev  (BSK) explicitly accounted for ion coupling based on a Landau-Ginzburg type functional for the free energy.\cite{BSK2011} Analysis of the Ornstein-Zernike equation for binary ionic fluids has also predicted non-monotonic screening lengths in concentrated electrolytes. \cite{Carvalho1994} Recently, the mean spherical approximation screening length\cite{Blum1977} has been used to study the ion current rectification in bipolar nanopores. \cite{Fertig2021} \\
\indent Local density approximations have also proven useful in capturing integrated quantities such as the capacitance of the diffuse layer near a charged surface \cite{ Squires2015, Khair2017} and important electrokinetic phenomena. \cite{Bazant2009,Khair2017} Stout and Khair \cite{Khair2017} used Bikerman\cite{Bikerman1942} and Carnahan-Starling\cite{Carnahan1969} type models to account for the entropic effects due to ion size to study the importance of steric interactions on diffusiophoresis in concentrated electrolytes. Nonetheless, these local density approximations are unable to capture the charge density oscillations present in concentrated electrolytes,\cite{Squires2015} do not show underscreening, and have been found to be ill-equiped to accurately describe the electric double layer structure. \cite{Gillespie2015,Khair2017}\\
\indent More recently, Adar et al. \cite{Andelman2019} modified the classical Coulomb potential to account for the finite volume over which an ion's charge is distributed, and examined its importance in setting the correlation length for charge-charge interactions. We refer to this model as the ``shell model'' throughout our analysis. The authors build a model for symmetric electrolytes and weak ion-ion electrostatic interactions with the ions represented by homogeneously charged spherical shells that are allowed to overlap.  The shells regularize the singularity in the standard Coulomb potential. From this model, they show that the Fourier transform of the charge-charge correlation function in the electrolyte is
\begin{equation}
   S(k)=\frac{ne^{2}k^{2}}{k^{2}+\kappa_{\mathrm{D}}^2 \hat{h}(2ka)}\;,  
   \label{eq:Andelam_S}
\end{equation} 
where ${n}$ is the number density of the ions, ${ a }$ is the shell radius and $ \hat{h}(x)=(2\sin(x/2)/x)^{2} $.  They extract the spatial decay of the charge-charge correlation from the generally complex-valued poles of $ {S (k)} $. The long length scale correlations appear at small ${k}$, and this behavior can be well described by the pole of $ S(k) $ whose imaginary part is closest to the real axis.  From analysis of this pole, they find that the exponential decay length in charge correlations, $ 1 / \kappa $, follows the Debye-H\"uckel predictions for low ionic strengths.  They also find that  ${\kappa a \sim 1/(\kappa_{\mathrm{D}} a)}$ as $ \kappa_{\mathrm{D}} a \rightarrow \infty$.  Here, the shell radius $ a $ establishes dimensional consistency in the scaling relation derived analytically at high ionic strengths.  The underscreening observed in this asymptotic limit of the charge correlation function is weaker than measured with surface forces experimentally.  Furthermore, this asymptotic power law scaling is only realized in their model when $ \kappa_{\mathrm{D}} a $ is $ O( 10^2 ) $ and greater.  Over a more realistic range of values, $ 1 < \kappa_{\mathrm{D}} a < 10 $, their model exhibits an effective power law scaling that is even weaker: $ \kappa a \sim 1/(\kappa_{\mathrm{D}} a)^p $ with $ p \approx 0.75 $. \\
\indent Additionally, Molecular Dynamics (MD) simulations have been used to study the bulk ionic screening length of concentrated electrolytes and ionic liquids.\cite{coles_rotenberg2020,zeman_holm2020} The screening length is extracted from the radial distribution function of the ions in bulk. Coles and colleagues \cite{coles_rotenberg2020} investigated ${\mathrm{LiCl}}$ and ${\mathrm{NaI}}$ in water and lithium bis(trifluoromethane)sulfonimide (${\mathrm{LiTFSI}}$) in water and an organic solvent. Their findings agree with the theoretical prediction of non-monotonic dependency of the inverse of the screening length, ${\kappa}$, on the concentration of ions. However, the scaling expression found in this limit, ${\kappa \sim 1/\kappa_{\mathrm{D}}^{0.3}}$, has a significantly weaker power than that measured experimentally or predicted from mean field modeling. Following a similar method to extract the correlation lengths, Zeman et al. \cite{zeman_holm2020} investigated 1-butyl-3-methylimidazolium hexafluorophosphate (${\mathrm{[C_{4}C_{1}Im]^{+}[PF_{6}]^{-}}}$) and ${\mathrm{NaCl}}$ in water. By fitting the radial distribution to the sum of two exponentially damped oscillations, the authors were able to obtain two correlation lengths. One of the correlation lengths scales linearly with ${\kappa_{\mathrm{D}}}$, whereas the second is estimated to scale as ${ \kappa \sim 1/\kappa_{\mathrm{D}}}$.
\\
\indent Furthermore, Cats et al. \cite{Cats_VanRoij2020} used Density Functional Theory (DFT) to study the electric double layer (EDL) and correlation lengths in a system comprised of a planar electrode and an aqueous electrolyte. The authors compare the results against those found by the corresponding two-body correlation function in bulk systems obtained from Integral equation Theory and MD simulations. In the MD simulations, the correlation length in the far field scales as ${ \kappa \sim 1/\kappa_{\mathrm{D}}^{0.3}}$. The authors use DFT, mean spherical approximation, and a mean field treatment of Coulombic interactions to obtain correlation lengths that range from ${ \kappa \sim 1/\kappa_{\mathrm{D}}^{0.1}}$ to ${ \kappa \sim 1/\kappa_{\mathrm{D}}^{0.6}}$. It is necessary to acknowledge that there is a limited range of $ \kappa_D $ over which these power law scaling relations can be extracted, and they should be considered descriptive rather than precise. \\
\indent Here, we use Brownian Dynamics to simulate a version of the Restrictive Primitive Model (RPM) and to study the importance of excluded volume interactions, ionic charge regularization, and ionic strength in determining concentrated electrolytes' structural properties. To study the competition of electrostatic and excluded volume interactions, we explicitly account for two physical length scales in our analysis. The first length scale is referred to as electrostatic radius or shell radius, ${a}$, over which the ion's charge is distributed. The second length scale,  ${a_{\mathrm{hs}}}$, arises from the hard core interactions between the ions in this model and is called the hard-sphere radius. One way to conceptualize these length scales is to imagine that the charged particles exist in a lattice in which the inter-particle distances are restricted by the hard core length scale $ a_\mathrm{hs} $, while the electrostatic interactions between ions are Coulombic for interparticle separations larger than ${a}$.  The hard sphere radius, ${a_{\mathrm{hs}}}$, determines an effective degree of ``swelling'' of the ions.  In the limit that $ a_\mathrm{hs} \gg a $, the ions are like point charges with inter-ion separations bigger than $ a_\mathrm{hs} $.  When $ a_\mathrm{hs} < a $, the ions are charged shells that can slightly overlap, but still have a solid (if small) core. By decoupling the length-scale over which the charge is regularized from the hard-sphere radius, we can compare our analysis against mean-field theories that consider ${(a_{\mathrm{hs}}/a)\rightarrow0}$. Specifically, we compared our results to those obtained by the mean-field model by Adar et al. in the limit that $ a_\mathrm{hs} \ll a $. Because real ions in solution are hydrated, the swelled (or deswelled) core and shell model is an interesting minimal representation of electrolytes extending the RPM.\\
\indent We extract the inverse correlation length, ${\kappa}$, and frequency of the charge density oscillations, ${\omega}$, using two different methods. First, we use the non-uniform Fourier transform of the ion centers to obtain the charge density structure factor ${S(\mathbf{k})}$ (as presented in equation \ref{eq:SofQ}). In this method, the values of ${\kappa}$ and ${\omega}$ are obtained from the small ${k}$ profile of the charge density structure factor where long length scale correlations appear. The obtained profiles for ${\kappa}$ and ${\omega}$ are then compared against those found by analyzing the decay of the long range charge density correlations in real space. In contrast to the method used to obtain the charge density structure factor, the charge density correlation function in real space is obtained using the well known ``shell summation'' method.\cite{kjellander_2001,Frenkel2002}
\\
\indent As simple theoretical models suggest that there is no particular preference for ions to reside in neutral pairs,\cite{lee_goriely2014} we study the spatial correlation between ions sharing the same sign in charge. In examining the correlations between like-charges, we find a positive correlation between ions with charges of equal sign at short distances and high concentrations. This observation points to the existence of complex clusters.  We perform a cluster analysis on like-charge ions in our model and compare the cluster size probability distribution against that of a binary hard sphere liquid, thereby revealing new insight into the structure of concentrated electrolytes in bulk.
\section{\label{sec:SimMethod} Methods}
\subsection{ Brownian Dynamics}
We model the ions in the electrolyte as a suspension of charged, hard, spherical particles. The solvent is treated implicitly and assumed to be Newtonian, such that it interacts with the ions through hydrodynamic forces from flows in the medium, and stochastic Brownian forces from fluctuating hydrodynamics due to momentum relaxation of the solvent molecules. \cite{Frenkel2002} On the ion scale, inertial relaxation occurs on time scales orders of magnitude smaller than those on which ions move. In this regime, any perturbation to ion momentum is felt almost instantaneously, and the ion moves at its terminal velocity, allowing us to neglect inertia. Under these assumptions, the overdamped Langevin equation governs the dynamics of the ions:
\begin{equation}
0=\mathbf{F}_{\alpha}^{\mathrm{H}}+\mathbf{F}_{\alpha}^{\mathrm{I}}+\mathbf{F}_{\alpha}^{\mathrm{E}}+\mathbf{F}_{\alpha}^{\mathrm{B}}\:,
\label{eq:overdamped}
\end{equation}
where $\mathbf{F}_{\alpha}^{\mathrm{H}}$ is the hydrodynamic force acting on the
$\alpha^{\mathrm{th}}$ ion, $\mathbf{F}_{\alpha}^{\mathrm{I}}$ accounts for forces arising from a generic conservative potential, $\mathbf{F}_{\alpha}^{\mathrm{E}}$ is the external force exerted by a global field \cite{Sherman2019} and $\mathbf{F}_{\alpha}^{\mathrm{B}}$ is the stochastic, Brownian force. The last force satisfies the fluctuation-dissipation theorem \cite{Russel1989} with ensemble average:
\begin{equation}
    {\left\langle \mathbf{F}^{\mathrm{B}}(t)\right\rangle =0};\hspace{4mm}\left\langle \mathbf{F}^{\mathrm{B}}(t)\mathbf{F}^{\mathrm{B}}(t+\tau)\right\rangle =2k_{\mathrm{B}}T(\mathbf{M}^{\mathrm{H}})^{-1}\delta(\tau),
\end{equation}
where $ \mathbf{F}^B(t) = [ \mathbf{F}^B_1(t), \mathbf{F}^B_2(t), \ldots ] $, $ \left\langle \cdot\right\rangle $ indicates the expectation value, $\delta$ is the Dirac delta function and $\mathbf{M}^{\mathrm{H}}$ is the hydrodynamic mobility tensor. This formulation ensures that any energy an ion gains from a thermal fluctuation is dissipated as drag to the solvent.\\
\indent The hydrodynamic mobility tensor couples the non-hydrodynamic force, $\mathbf{F}_{\beta}=\mathbf{F}_{\beta}^{\mathrm{I}}+\mathbf{F}_{\beta}^{\mathrm{E}}+\mathbf{F}_{\beta}^{\mathrm{B}}$, to the velocity of the $\alpha^{\mathrm{th}}$ ion:
\begin{equation}
   \mathbf{u}_{\alpha}(t)= \sum_{\beta=1}^{N}\mathbf{M}_{\alpha \beta}^{\mathrm{H}}\cdot\mathbf{F}_{\beta}(t)\;.
   \label{eq:velocity}
\end{equation}
\indent As we are interested in equilibrium properties, interparticle hydrodynamic interactions can be neglected. In this case, the drag on each ion is decoupled from all of the others and is equal to the Stokes drag,  
\begin{equation}
    \mathbf{M}_{\alpha \beta}^{\mathrm{H}}=0,\alpha \neq \beta ;\:\hspace{4mm}\mathbf{M}_{\alpha \alpha}^{\mathrm{H}}=\mathbf{I}/\gamma\;,
\end{equation}
where all ions are assigned the same drag coefficient, $ \gamma $.  Equation \ref{eq:overdamped} can be numerically solved via an Euler-Maruyama integration scheme:
\begin{equation}
\mathbf{x}_{\alpha}(t+\Delta t)=\mathbf{x}_{\alpha}(t)+ \mathbf{u}_{\alpha}(t)\Delta t\:,
\label{eq:forward}
\end{equation}
where $ \Delta t $ is the time step over which ion trajectories are advanced.\\
 \indent Forces arising from conservative interactions among ions are represented as the gradient of a potential energy $U(\mathcal{X})$, which is a function of the coordinates of all ions $\mathcal{X} \equiv [\mathbf{x}_1, \mathbf{x}_2, \dots, \mathbf{x}_N]^T$,
\begin{equation}
\mathbf{F}_{\alpha}^{\mathrm{I/E}}(\mathcal{X}) \equiv -\nabla_{\mathbf{x}_{\alpha}} U^{\mathrm{I/E}}(\mathcal{X}),
\label{eq:potential_grad}
\end{equation}
where the gradient is taken with respect to the position of the $\alpha$\textsuperscript{th} particle.  \\
\indent In our analysis, we are concerned with finite sized ions, which cannot overlap. The hard sphere force computed by the derivative of the well known step-like potential is discontinuous; it is zero everywhere except for a $\delta$-function of infinite magnitude at contact. This type of potential cannot be implemented directly in simulations.  Typically, the hard potential is approximated with a soft potential of the form $r^{-n}$, where $n$ is a large power.  The larger $n$ is, the more accurately the soft potential approximates the hard potential, but the resulting force becomes larger as the potential diverges increasingly quickly. \cite{Heyes1994A}  Smaller time steps must therefore be taken as $n$ increases to prevent unphysically large steric forces, rendering this method computationally inefficient.  Heyes and Melrose implemented a ``potential-free'' hard sphere algorithm by allowing particles to overlap over the course of a time step due to other forces and then separating them to contact at the end of the time step.\cite{Heyes1993} This implementation of the hard sphere potential allows us to use an integration time-step equal to ${10^{-3}\tau_{\mathrm{D}}}$.\\
\indent Because equations \eqref{eq:velocity} and \eqref{eq:forward} give a relation between particle displacements and forces, we can compute the effective force that is required to move two overlapping ions back into contact following one time step.  Thus, the potential-free algorithm can be equivalently written in terms of a hard sphere pair potential:\cite{Varga2015}
\begin{equation} \label{eq:hardsphereFD}
U_{\alpha \beta}^{\mathrm{hs}}(r)=\begin{cases}
\frac{\gamma}{4\Delta t}(r-2a_{\mathrm{hs}})^{2} & \text{if }r<2a_{\mathrm{hs}}\\
0 & \text{if }r\geq2a_{\mathrm{hs}}
\end{cases} \: ,
\end{equation}
where $a_{\mathrm{hs}}$ is the effective hard sphere radius of an ion, which can be different from the radius of the shell over which the ion's charge is distributed, ${a}$. In the form of the potential used here, ${\gamma}$ and ${a_{\mathrm{hs}}}$ are the same for all particles.  The total potential energy due to the hard sphere repulsion can be decomposed into a sum of pair potentials ${U_{\alpha \beta}}$
\begin{equation}
    U^{\mathrm{hs}}(\mathcal{X})=\frac{1}{2}\sum_{\alpha,\beta}U_{\alpha \beta}^{\mathrm{hs}}(|\mathbf{x}_{\alpha}-\mathbf{x}_{\beta}|)\: ,
\end{equation}
 where ${\alpha}$ and ${\beta}$ indices run over all particles, and the factor of ${1/2}$ corrects for double-counting each pair.\\
\indent Charged species feel additional electrostatic forces that can be obtained by solving Poisson's equation for the scalar potential,$\psi(\mathbf{x})$. \cite{Jackson1999, Landau1984} In this work, we regularize the potential near an ion by localizing its charge to a spherical shell of radius $ a $.  All the free charge is located on this surface such that Poisson's equation reduces to Laplace's equation for the potential outside and inside each ion shell:
\begin{equation}
    \nabla^2\psi=0\:,
    \label{eq:laplace}
\end{equation}
with boundary conditions on the shell of ion $ \alpha $ given by
\begin{equation}
    \psi_{\mathrm{p}} = \psi_{\mathrm{f}} , \hspace{4mm} (\mathbf{E}_{\mathrm{f}} - \mathbf{E}_{\mathrm{p}}) \cdot \hat{\mathbf{n}} = q_{\alpha}/(4 \pi a^2)\:, \hspace{2mm}
    \label{eq:BCs}
\end{equation}
where ${\psi_{\mathrm{p}}}$ and ${\psi_{\mathrm{f}}}$ are the potentials inside and outside of the shell, ${q_{\alpha}}/(4\pi a^2)$ is the uniform free surface charge density of ion $ \alpha $ on a spherical shell of radius $ a $ and net charge ${q_{\alpha}}$,  $\mathbf{\hat{n}}$ is the normal outward vector, and ${\mathbf{E}_{\mathrm{f}} }$ and ${\mathbf{E}_{\mathrm{p}} }$ are the electric field outside and inside of the particle respectively. \cite{Jackson1999, Landau1984} As ions move with time or any external field varies, the boundary conditions presented in equation \ref{eq:BCs} vary with time. Thus, time-dependence in Laplace's equation emerges solely through the time-varying boundary conditions, and the electric potential equations are said to be pseudo-steady. \cite{Sherman2018}\\
\indent For a point ${\mathbf{x}}$ in the fluid, the potential is given by the integral form of Laplace’s equation: \cite{Sherman2019}
\begin{align}
   \psi_{\mathrm{f}}(\mathbf{x})-\psi_{0}(\mathbf{x})=\\
  && \frac{1}{\varepsilon_{\mathrm{f}}}\sum_{\alpha}\int_{S_{\alpha}}d\mathbf{x}'(G(\mathbf{x}-\mathbf{x}')\mathbf{E}_{\mathrm{f}}(\mathbf{x}')\cdot\hat{\mathbf{n}}_{\mathbf{x}'} \nonumber \\
   &&+\varepsilon_{\mathrm{f}}\psi_{\mathrm{f}}(\mathbf{x}')\hat{\mathbf{n}}_{\mathbf{x}'}\cdot\nabla_{\mathbf{x}'}G(\mathbf{x}-\mathbf{x}'))\:,\nonumber
\end{align}
where ${\psi_{0}}$ is an externally imposed potential, $ S_{\alpha} $ is the spherical shell around ion $ \alpha $, $ \mathbf{\hat n} $ is the normal to the surface, $ \epsilon_{\mathrm{f}} $ is the permeability of the fluid, and, for a model of ions in the bulk, ${G}$ is a periodic Green's function in three dimensions. As the periodic Green's function is a solution to Laplace's equation, we can perform a multipole expansion to calculate the solution for a generic spherical particle in terms of the moments of the fluid potential. In the primitive electrolyte model, the first harmonic in the expansion accounts for perturbations to the potential field due to the charge.  We truncate at that level so that the fluid potential in a cubic simulation box with volume $ V = L^3 $, is:  
\begin{equation}
    \psi_{\mathrm{f}}(\mathbf{x})-\psi_{0}(\mathbf{x})=\frac{1}{\varepsilon_{\mathrm{f}}V}\sum_{\mathbf{k}\neq0}\sum_{\alpha}\frac{e^{i\mathbf{k}\cdot(\mathbf{x}-\mathbf{x}_{\alpha})}}{k^{2}}q_{\alpha}j_{0}(ka)\;,
    \label{eq:psi_f}
\end{equation}
where ${j_{0}(x)=\sin(x)/x}$ is the spherical Bessel function of the first kind, ${\mathbf{k}\in\left\{ (2\pi k_{1}/L,2\pi k_{2}/L,2\pi k_{3}/L):(k_{1},k_{2},k_{3})\in\mathbb{Z}\right\} }$, and $ k $ is the magnitude of $\mathbf{k}$.  Removing the ${\mathbf{k}=0}$ term from the wave space sum reflects electroneutrality of the electrolyte.\cite{Jackson1999}\\
\indent As there is no charge inside the particles, the potential ${\psi_{\mathrm{p}}}$ must also satisfy Laplace's equation and be expanded about the particle using spherical harmonics. The boundary conditions of this problem make it possible to relate the particle and fluid moments. This allows us to construct a system of equations by equating ${\psi_{\mathrm{f}}}$ and ${\psi_{\mathrm{p}}}$ at the surface of a particle and integrating over its surface:
\begin{equation} \label{eq:Moments}
\left\langle \Psi\right\rangle -\Psi_{0}=\mathcal{M}^{E}_{ \Psi q}\cdot\mathcal{Q}
\end{equation}
where on the left hand side of the equation $\left\langle \Psi\right\rangle -\Psi_{0}=[\left\langle \psi_{1}\right\rangle -\psi_{0}(\mathbf{x}_{1}),\left\langle \psi_{2}\right\rangle -\psi_{0}(\mathbf{x}_{2}),...,\left\langle \psi_{N}\right\rangle -\psi_{0}(\mathbf{x}_{N})]^{T}$
is a list of the relative potentials for each of the ions, $\psi_{0}$
being the externally imposed potential field at the ion center and $\left\langle \psi _{\alpha}\right\rangle$ being the surface averaged potential of the ${\alpha^{\mathrm{th}}}$ ion. On the right hand side, $ \mathcal{Q} =[q_{1},q_{2},...,q_{N}]$ is the list of ion charges, and $\mathcal{M}_{\Psi q}^{E}$
is known as the potential matrix, as it relates the zeroth moment of the surface charge distribution on the ions to the surface averaged potential. \cite{Sherman2019} \\
\indent The potential-charge coupling in $\mathcal{M}_{\Psi q}^{E}$ between ions $ \alpha $ and $ \beta $ is given by:
\begin{equation}
\frac{1}{\varepsilon_{\mathrm{f}}V}\sum_{\mathbf{k}\neq0}j_{0}^{2}(ka)\frac{e^{i\mathbf{k}\cdot(\mathbf{x}_{\alpha}-\mathbf{x}_{\beta})}}{k^{2}},
\label{eq:M00_Fourier}
\end{equation}
\\
and is a Coulomb-type interaction that is regularized when ions' centers are closer than $ 2a $ so that their charged shells overlap.  Evaluation of matrix elements like those in equation \ref{eq:M00_Fourier} is computationally expensive as the summand decays algebraically as ${1/k^4}$, and requires a large number of summed terms to converge. To improve the efficiency of the computations, we use a matrix-free method to compute the ion potentials, and accelerate the summations by introducing an Ewald splitting function, ${h(k)\equiv\exp(-k^2 / 4\xi^2 )}$. The function allows us to split the summation into two rapidly convergent series, one in real space, and the other in wave space:\cite{Sherman2019,Lindbo2011} 
\begin{align}
   && \left\langle \psi\right\rangle _{\alpha}-\psi_{0}(\mathbf{x}_{\alpha})=\frac{\varepsilon_{\mathrm{f}}}{V}\sum_{\mathbf{k}\neq0}\sum_{\beta}\frac{e^{i\mathbf{k}\cdot(\mathbf{x}_{\alpha}-\mathbf{x}_{\beta})}}{k^{2}}h(k)j_{0}^{2}(ka)\cdot q_{\beta} \nonumber \\
    &&+\varepsilon_{\mathrm{f}}\sum_{\mathbf{n}}\sum_{\beta}\mathcal{F}^{-1}\left\{ \frac{e^{i\mathbf{k}\cdot(\mathbf{x}_{\alpha}-\mathbf{x}_{\beta})}}{k^{2}}\left(1-h(k)\right)j_{0}^{2}(ka)\right\} \cdot q_{\beta},
    \label{eq:splitting}
\end{align}
where ${\mathcal{F}^{-1}}$ is the inverse Fourier transform. The Ewald splitting parameter, ${\xi}$, controls the rate of convergence of the real space and wave space sums.  Large values of ${\xi}$ increase the convergence speed of the real space sum, and small values increase the convergence of the wave space sum.\cite{Lindbo2011}  The second term on the right-hand side of equation \ref{eq:splitting} is obtained using the Poisson summation formula. The sum over all periodic images ${\mathbf{n}}$ can be computed pairwise using the inverse transform found in appendix A of Sherman (2019).\cite{Sherman2019} The wave space sum is calculated using the spectral Ewald method as detailed in references. \cite{Lindbo2011,Sherman2019}  With this approach, the ion potentials can be determined from their charges with log-linear computational complexity, enabling rapid simulations of up to $ O(10^5) $ ions in this work. \\
\indent The total electric potential energy of the electrolyte is:
\begin{equation}
    U^{\mathrm{E}}=\frac{1}{2}\mathcal{Q}\cdot(\left\langle \Psi\right\rangle -\Psi_{0}).
    \label{eq:E_potential}
\end{equation}
Using equation \ref{eq:Moments} and taking the gradient of the electric potential energy, we can calculate the electric force on the ${\alpha^{\mathrm{th}}}$ particle:
\begin{equation}
    \mathbf{F}_{\alpha}^{\mathrm{E}}=-\sum_{\beta}\nabla_{\mathbf{x}_{\beta}}\mathcal{\mathcal{M}}_{\psi q}^{E}q_{\alpha}q_{\beta}.
\end{equation}
This force can be computed using the same matrix-free methods as the potential.\cite{Sherman2019} Real electrolytes might have ions that interact with more complicated potentials due to the structure and fluctuations of hydration shells around the ions.  We have aimed here to examine the simplest possible model where the effects of other structural features are captured solely by the hard core repulsion.\\
\indent All simulations are of binary and symmetric electrolytes; thus, the charge of the ions of different species is of equal magnitude but opposite sign. The simulations are made dimensionless by measuring distances relative to ${a}$, time in units of the ion diffusion time: ${\tau_{\mathrm{D}}=k_{\mathrm{B}}T/\gamma a^2}$, and charge relative to the charge scale: $\sqrt{\varepsilon_{\mathrm{f}} a k_{\mathrm{B}} T} $. We introduce an electric volume fraction, ${\phi_a}$, calculated with respect to the electric radius $ a $ instead of ${a_{\mathrm{hs}}}$, and we use the subscript ${a}$ to highlight this distinction. The length of the simulation box ${L}$ is the same in all simulations: ${100a}$. The investigated electric volume fraction of ions (cations + anions) varies from ${\phi_a=0.001}$ to ${\phi_a=0.45}$ (approximately ${ 5 \times 10^{-3} \: \mathrm{mM} }$ and $ {2.5 \: \mathrm{M} }$ solutions of ${ \mathrm{NaCl} }$ in water, respectively, assuming that ${ a= a_{ \mathrm{hs} } = 0.33 \: \mathrm{ nm }}$ on average). The strength of the charge-charge interactions is prescribed by the dimensionless parameter $\epsilon$ (not to be confused with the permittivity of the solvent $\varepsilon_{\mathrm{f}}$). We define this quantity as the electric potential between two point charges at a distance of ${2a}$ with respect to the thermal energy: 
\begin{equation}
   \epsilon=\frac{q_{i}^2}{8 \pi \varepsilon_{\mathrm{f}}a k_\mathrm{B} T} = \frac{\lambda_{\mathrm{B}}}{2a}  \;.
\end{equation} 
\indent The studied values of the electrostatic interactions relative to ${ k_{\mathrm{B}}T }$ are chosen to be within the range of the values frequently found in common electrolytes: ${\varepsilon = [0.5, 2.0, 5.0] }$. For a solution of ${\mathrm{NaCl}}$ in water at room temperature, ${\varepsilon\approx1.1}$ (assuming relative permittivity of $80$, and ${a\approx 0.33 \times 10^{-9}\:\mathrm{m}}$); if methyl formaldehyde is used as the solvent, ${\varepsilon\approx0.47}$ (assuming relative permittivity of $182.4$). Higher values of ${\varepsilon}$, at room temperature, are more commonly found in non-aqueous electrolytes. For example, for ${\mathrm{LiCl}}$ in ethanol, $\varepsilon \approx 4.6 $  (assuming ${a\approx 0.25 \:\mathrm{nm}}$, and relative permittivity of ${24.3}$).
The radius of the hard sphere interactions, ${a_{\mathrm{hs}}}$, ranges from ${0.5a}$ to ${1.25a}$.  This range of parameters allows us to investigate the competitive effects of excluded volume and electrostatic interactions in the simulation. The simulations were performed in HOOMD-Blue.\cite{HOOMD} Simulations where $200\tau_{\mathrm{D}}$ long, and the first $100\tau_{\mathrm{D}}$ were discarded to guarantee equilibrium. The integration step is set to ${10^{-3}\tau_{\mathrm{D}}}$. A detailed description about the equilibration process is found in Appendix \ref{sec:Relaxation}. 

\begin{figure*}[ht!]
\includegraphics[width=0.98\textwidth]{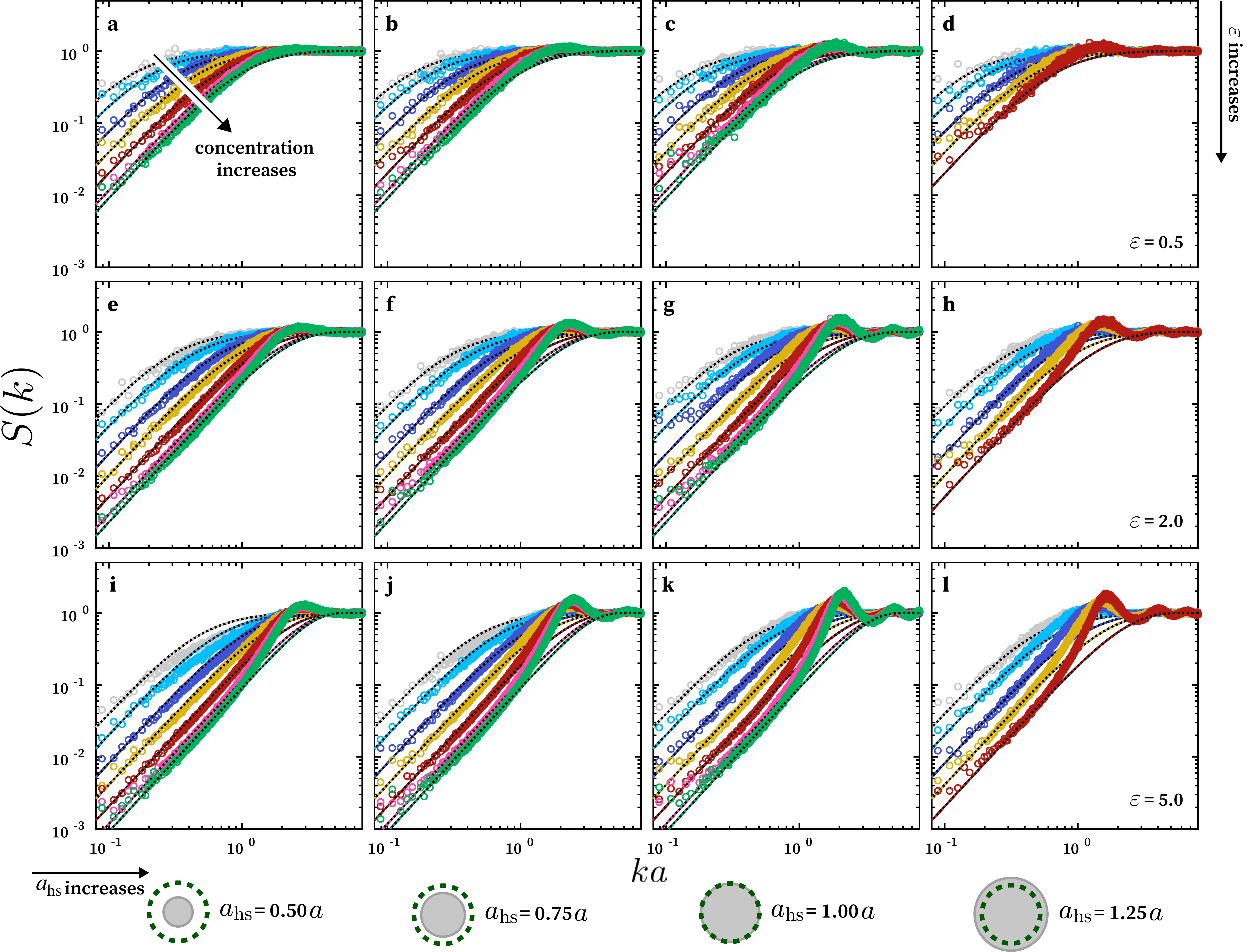}
\caption{\label{fig:SofQ} Evolution of the charge structure factor upon change in concentration of ions, hard sphere radius, and strength of the ion-ion interactions. Hard sphere radius increases from ${a_{\mathrm{hs}}=0.50a}$ (first column) to ${a_{\mathrm{hs}}=1.25a}$ (fourth column). The strength of the ion-ion interactions increases moving down from ${\epsilon = 0.5}$ (first row, subfigures $\mathbf{a}$-$\mathbf{d}$) to ${\epsilon = 5.0}$  (third rown, subfigures $\mathbf{i}$-$\mathbf{j}$). Lines are obtained using the theory of Adar et al.\cite{Andelman2019} The model is in close agreement with the simulations at low concentrations of ions, and particularly in the cases where ${a_{\mathrm{hs}}<a}$ (first and second columns). Nonetheless, it fails to predict the structural features that arise as the concentration increases -- such as the increase in the curvature in the vicinity of ${ka=2}$.  In each sub-figure, the ion concentration grows from top to bottom with ion shell volume fractions of $ \phi_a =[ 0.01,0.02,0.05,0.10,0.20,0.35,0.45] $ represented.  For the largest hard-core ions (fourth column), the case $ \phi_a =0.35 $ and $ \phi_a =0.45 $  are absent as they give a hard core volume fraction that exceeds the random close-packing limit.}
\end{figure*}
\section{\label{sec:Results} Analysis and Discussion}
\subsection{\label{sec:SofQ_Methods} Charge Density Correlations}
\indent The spatial distribution of charge in an electrolyte can be represented in terms of the charge density:
\begin{equation}
\rho( \mathbf{x} ) = e \sum_{\alpha=1}^N z_{\alpha} \delta( \mathbf{x} - \mathbf{x}_{\alpha} ),
\end{equation}
where index $ \alpha $ runs over all $ N $ ions in the volume of electrolyte $ V $, $ e $ is the fundamental charge, $ z_{\alpha} $ is the valence of ion $ \alpha $, and $ \mathbf{x}_{\alpha} $ is the position of ion $ \alpha $. The Fourier transformation of the charge density distribution is:
\begin{equation}
    \hat \rho( \mathbf{k} ) = \int \exp(-i \mathbf{k} \cdot \mathbf{x} ) \rho( \mathbf{x} ) \, d \mathbf{x} = e \sum_{\alpha=1}^N z_{\alpha} \exp( -i \mathbf{k} \cdot \mathbf{x}_{\alpha}),
\end{equation}
which can be used to define a charge structure factor:
\begin{equation}
S( \mathbf{k} ) = \frac{\left< \hat \rho( \mathbf{k} ) \hat \rho^*( \mathbf{k} ) \right> - \left< \hat \rho( \mathbf{k} ) \right> \left< \hat \rho^*( \mathbf{k} ) \right>}{V e^2 \sum_{\nu=1}^M n_\nu z_\nu ^ 2},
\label{eq:SofQ}
\end{equation}
where $\hat \rho^*( \mathbf{k} )$ is the complex conjugate of $\hat \rho( \mathbf{k} )$, the index $ \nu $ runs over all $ M $ different ion species, and $ n_{\nu} $ is the bulk number density of ions of that species.  For a $z$:$z$ electrolyte, the denominator reduces to: $ V n ( z e )^2 $, where $ n $ is the bulk number density of all ions. \\
\indent When the charge density fluctuations are normalized in this way, we can guarantee that $ S( \mathbf{k} ) \rightarrow 1 $ as $ k \rightarrow \infty $. Additionally, if charge neutrality is ensured: $ \sum_{\nu=1}^M n_\nu z_\nu = 0 $, then $ S( \mathbf{k} ) \rightarrow 0 $ as $ k \rightarrow 0 $.  For a $z$:$z$ electrolyte, the Stillinger-Lovett second moment condition also requires that: $ S( \mathbf{k} ) / k ^ 2 \rightarrow 2 / \kappa_{\mathrm{D}}^{2} $ as ${ k \rightarrow 0} $.\cite{Stillinger1968}\\
\begin{figure*}[ht!]
\includegraphics[width=0.9\textwidth]{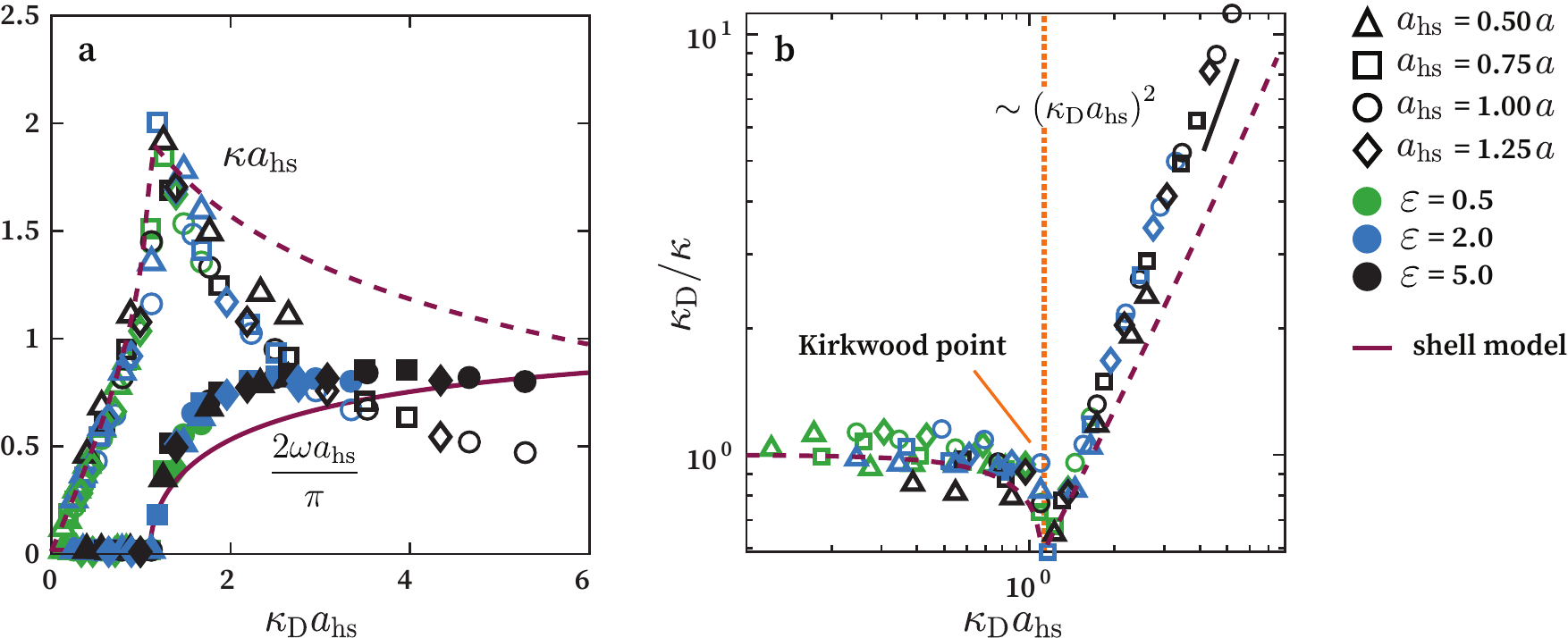}
\caption{\label{fig:decay_sofq} ($\mathbf{a}$) Extracted inverse decay length ${\kappa}$ (open markers) and oscillation wave number ${\omega}$ (filled markers) as a function of ${\kappa_{\mathrm{D}}a_{\mathrm{hs}}}$. These values were obtained from the charge structure factors, $S(k)$ (shown in Figure \ref{fig:SofQ}) by computing the complex root closest to the real axis of equation \ref{eq:polyn} as described in section \ref{sec:SofQ_Analysis}. Continuous and dashed lines represent the expected values for ${\omega}$ and ${\kappa}$ obtained from the poles of equation \ref{eq:Andelam_S}. ($\mathbf{b}$) The ratio of the charge-charge decay length to the Debye length as a function of $ \kappa_{\mathrm{D}}a_{\mathrm{hs}} $. The obtained scaling shows more underscreening than the scaling proposed by Adar et al. \cite{Andelman2019}, see section C.  The dotted orange line indicates where crossover from monotonic to oscillatory decay occurs.}
\end{figure*}
\indent The partial pair distribution function for a binary system is given by:\cite{hansen_mcdonald_2006}
\begin{equation}
    g_{\nu\mu}(\mathbf{r})=\frac{1}{Vn_{\nu}n_{\mu}}\left\langle \sum_{\alpha=1}^{N_{\nu}}\sum_{\beta \neq \alpha}^{N_{\mu}}\delta(\mathbf{r}+\mathbf{r}_{\alpha}-\mathbf{r}_{\beta})\right\rangle\:,
    \label{eq:binary_gofr}
\end{equation}
where ${\mathbf{r}_{\alpha/\beta}}$ is the position of particle ${\alpha}$ of species ${\nu}$ or of particle ${\beta}$ of species ${\mu}$.  In a single component system, particles ${\alpha}$ and ${\beta}$ are by necessity of the same species; in a mixture, we can define the pair distribution function with ${\nu}$ and ${\mu}$  representing the same or different species. The partial pair correlation function for distinctly charged regions in the electrolyte, $ h_{\nu\mu}(\mathbf{r}) = g_{\nu\mu}(\mathbf{r}) -1 $, is related to the charge-charge structure factor through the Fourier transformation. For an isotropic electrolyte, the charge correlations show no angular dependence, and thus $ g_{\nu\mu}( \mathbf{r} ) $ and $ h_{\nu\mu}( \mathbf{r} ) $ depend only on $ r = | \mathbf{r} | $. Similarly, $ S( \mathbf{k} ) $ depends only on $ k = | \mathbf{k} | $ so that:
\begin{equation}
  S(k)=\sum_{\nu,\mu}z_{\nu}z_{\mu}\left(\frac{n_{\nu}}{n}\delta_{\nu\mu}+4\pi\frac{n_{\nu}}{N}n_{\mu}\int_{0}^{\infty}\frac{\sin(kr)}{kr}h_{\nu\mu}(r)r^{2}dr\right),
  \label{eq:SofQfromGofR}
\end{equation}
where ${\delta_{\nu\mu}}$ is the Kronecker delta, which equals ${1}$ if ${\nu=\mu}$, and ${0}$ otherwise.
\subsection{ \label{sec:SofQ_Analysis}  Correlation Lengths Computed From Charge Density Structure Factor}
\indent We obtain the static structure factor for the charge density using equation \ref{eq:SofQ}. The Non-uniform Fast Fourier Transform (NUFFT) of the ion centers is computed using the library FINUFFT.\cite{FINUFFT} The minimum wave vector in each spatial dimension is ${2\pi/L}$, where ${L}$ is the length of the simulation box. Additionally, as we are interested in the small ${k}$ behavior, we only use the first ${32}$ Fourier modes, thus minimizing computational time without compromising accuracy for the studied range of ${\mathbf{k}}$. The structure factor is averaged over ${100}$ independently sampled configurations.\\
\begin{figure*}
\includegraphics[width=0.95\textwidth]{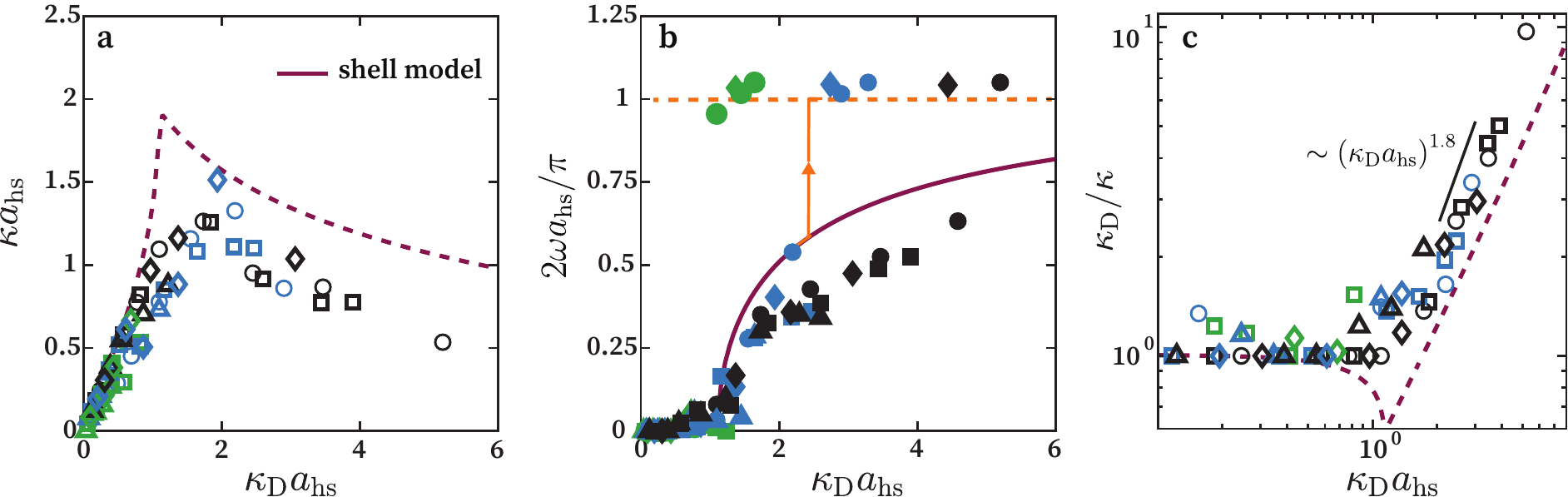}
\caption{\label{fig:kappa_omega_gofr} Extracted values for the inverse of the correlation length ${\kappa}$, ($\mathbf{a}$), and oscillation frequency ${\omega}$, ($\mathbf{b}$). These quantities were obtained from the charge-charge correlation function, ${h_{+-}(r)=g_{+-}(r)-1}$, following the optimization steps described in section \ref{sec:gofr_analysis}. The charge-charge pair distribution function, $g_{+-}(r)$ was obtained with the standard shell summation method. Note the sudden jump in the charge oscillation frequency in ($\mathbf{b}$); the jump is present at high concentrations when (even at the lowest strength) the inter-particle distance is less than the hard sphere diameter, forcing the system to order itself in discrete layers of charge delimited by the hard sphere length-scale as shown in Figure \ref{fig:gofr}. ($\mathbf{c}$) The charge-charge decay length as a function of the inverse of the Debye length. The obtained scaling is lower than that obtained from the charge structure factor ${S(k)}$, shown in Figure \ref{fig:decay_sofq} ($\mathbf{b}$).  The symbols follow from the legend in Figure \ref{fig:decay_sofq}.}
\end{figure*}
\indent Figure \ref{fig:SofQ} shows the charge density structure factor as a function of the concentration, strength of the ion-ion interactions (${\varepsilon}$), and hard sphere radius (${a_{\mathrm{hs}}}$). The dotted lines are obtained using the mean-field model in equation \ref{eq:Andelam_S}, which is valid in the limits that $ a_\mathrm{hs} \ll a $ and $ \epsilon \ll 1 $. At relatively low concentrations and small $k$, the simulation closely follows the behavior proposed by Adar et al., \cite{Andelman2019} particularly for ${a_{\mathrm{hs}}=0.50a}$ and ${\epsilon=0.50}$ (Figure \ref{fig:SofQ} ($\mathbf{a}$)). However, as the concentration, the value of ${a_{\mathrm{hs}}}$, and the strength of the ion-ion interactions increase, the model is unable to capture the behavior near ${ka=1}$. This is likely because the model in equation \ref{eq:Andelam_S} does not directly account for excluded volume interactions. It regularizes the electric potential when ions are closer than a diameter, and, as the hard sphere radius increases and packing effects become significant, additional structural features arise as a direct result of excluded volume interactions. Deviations from Stillinger-Lovett's second moment condition as ${k \rightarrow 0}$ are revealed by the charge structure factor. Importantly, there is an increase in the curvature of the values of ${S(k)}$ as ${ka}$ approaches unity from the left, and formation of additional local maxima in the simulated results.  These peaks look like the nearest neighbor peak and its reflections in a concentrated hard particle liquid, and such features are not realized by the shell model. \\
\indent The real space charge correlations, ${ h(r) }$,  can also be obtained via the inverse Fourier transformation of the charge density structure factor. When an analytical expression for ${S(k)}$ is known, this transformation can be evaluated by contour integration in the plane of complex wavenumbers, ${k=\omega + i \kappa }$. The value of the integral is the sum of the residues, ${R_{n}}$, at the poles:
\begin{equation}
    h(r) = \frac{ 1 }{ 2\pi }\sum_{n}R_{n}\frac{\exp{(k_{n} r)}}{r} \:.
    \label{eq:Residueh}
\end{equation}
The poles may lie on the imaginary axis ($\omega=0$) or may form a conjugate pair. In the first case, the contribution of the decay of ${rh(r)}$ is purely exponential; in the second case, there is a damped oscillatory contribution from the conjugate pair. In principle, there could be an infinite number of such terms but the presence of the exponential factors in equation \ref{eq:Residueh} ensures that asymptotically, the dominant contribution will come from the poles nearest to the real axis.
\indent The structure factor for the charge density obtained from simulations or experimental tools is measured only for real values of ${k}$. In contrast, analytical expressions such as equations \ref{eq:Andelam_S} and \ref{eq:SofQ} are valid for any complex ${k}$. This inability to obtain the full mapping of ${S(k)}$ presents a challenge in understanding the role of the poles and, consequently, extracting correlation lengths. Nonetheless, some characteristics of the structure factor in the complex plane can still be observed in ${S(k)}$ for real ${k}$. Since ${S(k)-1}$ is a holomorphic function of the complex variable ${k}$,\cite{Copson1984, Cummings1983} the Cauchy-Riemann equations imply that a maximum or minimum in the plane ${\mathrm{Im}(k)=0}$ corresponds to a saddle point in the complex ${k}$-plane. Therefore, the peaks we observe for real-valued ${k}$ as in Figure \ref{fig:SofQ} are saddle points in the complex plane. \cite{Cummings1983} \\
\indent To extract the values of ${\kappa}$ and ${\omega}$ from the Fourier transform of the charge-charge correlation function, we assume that ${S (k)}$ can be represented as a rational function:
\begin{equation}
    S(k)=\frac{k^{2}}{f(k)}.
    \label{eq:SofQf}
\end{equation}
As ${S(k)}$ is an even function of ${k}$, we propose ${f(k)}$ to have the following form at small $ k $:
\begin{equation}
    f(k)=\frac{\kappa_{\mathrm{D} }^{2}}{2}+\lambda_{2}k^{2}+\lambda_{4}k^{4}+\lambda_{6}k^{6}+\lambda_{8}k^{8}+\mathcal{O}(k^{10})\:,
    \label{eq:polyn}
\end{equation}
where the first term on the right-hand side of the equation immediately satisfies Stillinger-Lovett's second moment condition. Additionally, the roots of ${f(k)=0}$ correspond to the poles of equation \ref{eq:SofQf}.\\
\indent To numerically extract the small ${k}$ behavior, we find the values of the coefficients ${\lambda_n}$, $ n = 2, 4, 6, 8 $, using linear least-squares regression of $ k^2 / S( k ) $, followed by a nonlinear least-squares regression of $ \log S( k ) $ using the linear regression results as the initial guess. Then, we compute the complex roots of $ f( k ) $, which is straightforward because it is a fourth-order polynomial in ${k^2}$. The frequency of the charge density oscillation and the inverse of the correlation decay length are the real and imaginary parts of the pole, respectively.  We chose the pole closest to the real axis, which reflects the slowest rate of spatial decay. \\
\indent Figure \ref{fig:decay_sofq} (a) shows the extracted values of ${\omega}$ and ${\kappa}$ as a function of the inverse of the decay length ${\kappa_{\mathrm{D}}}$ scaled on the hard sphere radius. The continuous and dashed lines are obtained by solving for the zeros in the denominator of equation \ref{eq:Andelam_S}. While ${\kappa}$ follows very closely the values obtained using the shell model at low values of ${\kappa_{\mathrm{D}}a_{\mathrm{hs}}}$, it decays more rapidly as ${\kappa_{\mathrm{D}}a_{\mathrm{hs}}}$ increases. The same is true for the frequency of oscillations, ${\omega}$, extracted from the charge structure factor; ${\omega}$ increases in value more rapidly with ${\kappa_{\mathrm{D}}a_{\mathrm{hs}}}$ than those values predicted by Adar et al. We believe this is likely due to the additional structural features of ${S(k)}$ sampled in the simulations. The difference in the renormalization of the ions' size deserves mentioning. While the length scale used in the mean field theory is the ionic radius, we found that the scale of decay is set by the hard-sphere radius, underscoring again the importance of excluded volume interactions in setting the correlation length.  \\
\indent The ratio ${\kappa_{\mathrm{D}}/\kappa}$ as a function of the Debye screening length, scaled on the hard core size, is shown in Figure \ref{fig:decay_sofq}(b) (right). At weak ionic strengths, the values of $\kappa$ scaled on the hard core size are in good agreement with the model predictions of Adar et al. In contrast, for high values of ${\kappa_{\mathrm{D}}}$, the amount of measured underscreening is stronger than previous simulations. For values of ${\kappa_{\mathrm{D}}}$ above the Kirkwood point, Cats and coworkers\cite{Cats_VanRoij2020} find that ${ \kappa \sim 1 / \kappa_{\mathrm{D}}^{0.3}} $ and Coles and colleagues\cite{coles_rotenberg2020} report that $ \kappa \sim 1 / \kappa_D^p$ with p < 0.5.  The decay lengths they extract are more scattered than in other simulation studies. Slightly stronger underscreening was obtained in DFT calculations, where it was found that ${ \kappa \sim 1 / \kappa_{\mathrm{D}}^{0.6}} $, approximately .\cite{Cats_VanRoij2020} Even stronger underscreening is identified in our simulations and analysis of this version of the RPM.  We find that ${ \kappa \sim 1 / \kappa_{\mathrm{D}}} $, with $ a_\mathrm{hs} $ used as the length scale to establish dimensional consistency in the scaling relation. The difference between the scalings predicted by different simulations could arise from a variety and combination of possible sources.  One possibility is that the charge structure factor used here is better suited for identifying long-ranged decay than the radial distribution function used in other studies. We test this idea in the next section.  Other possibilities, which are harder to test, include differences in the force fields used leading to different electrolyte microstructures and finite system size effects.  As with the mean-field model, charge oscillations appear at a finite value of $ \kappa_{\mathrm{D}} $ in the simulations -- a Kirkwood point located at $ \kappa_{\mathrm{D}} a_\mathrm{hs} \approx 1.1 $.
\subsection{\label{sec:gofr_analysis}Correlation Lengths Computed From Charge Density Pair Distribution Function}
\begin{figure}
\includegraphics[width=0.48\textwidth]{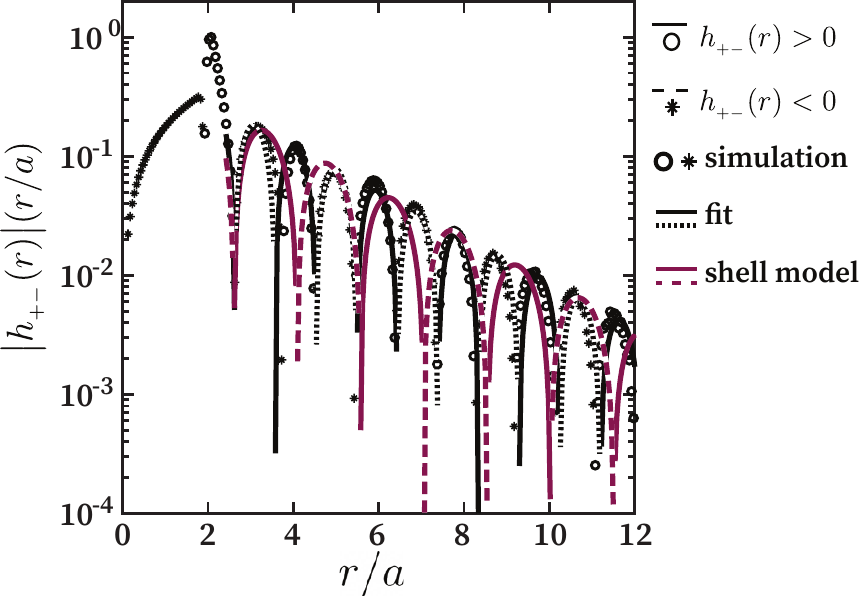}
\caption{\label{fig:gofr} Charge-charge correlation function for a symmetric binary electrolyte with ${\epsilon=2.0}$, $a_{\mathrm{hs}}=1.00a$, ${\phi =\phi_{a}= 0.45 }$, and ${\kappa_{\mathrm{D}} a_\mathrm{hs}=3.3}$. At this concentration, the screening length is on the order of the hard core diameter of the ions, and this ion size also dictates the frequency of charge oscillation. The oscillation period dictated by the shell model (pink line) is greater than the effective inter-particle distance. This results in the sudden jump shown in Figure \ref{fig:kappa_omega_gofr} ($\mathbf{b}$), where the oscillation period is dictated by the hard-sphere radius. }
\end{figure}
\begin{figure}
\includegraphics[width=0.48\textwidth]{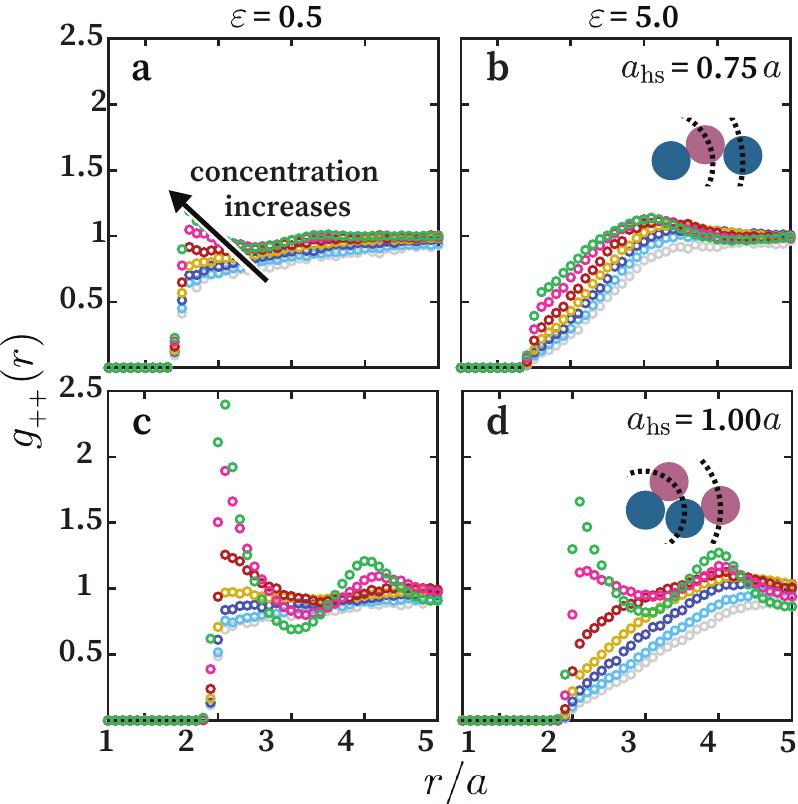}
\caption{\label{fig:gqq} Averaged self correlation function for equally charged species. In each sub-figure, the ion concentration decreases from top to bottom with ion shell volume-fractions of $ \phi_a =[ 0.01,0.02,0.05,0.10,0.20,0.35,0.45] $ represented. 
The strength of the ion-ion interactions, ${\epsilon=0.5}$ and  ${\epsilon=5.0}$ for the first column ($\mathbf{a}$ and $\mathbf{c}$) and for the second column ($\mathbf{b}$ and $\mathbf{d}$) respectively. The hard sphere radius is ${a_{\mathrm{hs}}=0.75a}$ and ${a_{\mathrm{hs}}=1.00a}$ for the first and second row respectively. Notice how the correlation near the first shell is less than one at low concentrations, and increases until it is greater than one. This behavior is highlighted as the hard sphere radius increases and ionic strength decreases.}
\end{figure}
\indent We sample the charge-charge correlation function using the standard ``shell summation".\cite{kjellander_2001} This method divides the space around a central particle into discrete shells and constructs a histogram by obtaining the average charge in each of the shells during the simulation. The histogram is normalized by each shell's volume.\\
\indent Just as a correlation length and oscillation frequencies can be extracted from the behavior of ${S(k)}$ as ${ k \rightarrow 0 }$, we can compute analogous quantities from ${ g_{ +- }(r) }$ by analyzing its large ${ r }$ behavior. We expect that for large values of ${ r }$, the pair distribution function will have the form:\cite{Carvalho1994}
 \begin{equation}
      h_{ +- }(r) = g_{ +- }( r ) - 1 \approx \frac{ A \exp( - \kappa r )}{ r } \cos( \omega r +\varphi) \; ,
      \label{eq:hr}
 \end{equation}
 where ${ A }$ is the amplitude, and ${ \varphi }$ is the phase angle. As equation \ref{eq:hr} is highly nonlinear, we start by extracting ${ \omega }$ by solving the optimization problem:
\begin{equation}
  \arg \min_{ (\omega,\varphi) }\sum_i \left[\sgn\left(\cos(\omega r_i+\varphi\right))\sgn\left(h_{+-}(r_i)r_i\right)\right],
\label{eq:sign}  
\end{equation}
 subject to $ \varphi\in[0,2\pi] $ to estimate ${\omega}$ and ${\varphi}$.  Here $ r_i $ are the centers of the different shells used to construct $ g_{+-}(r) $ via the shell summation method. \\
 \indent At a particular ionic strength, charge oscillations spontaneously emerge.  It is challenging to accurately infer $\omega$ in the neighborhood of the Kirkwood line.\cite{Kirkwood1954}  At the onset of charge oscillations, the oscillation frequency is vanishingly small, and observation of $ g_{+-}(r) $ over large distances is necessary to observe a complete cycle of oscillation. To generate a reliable value of ${\omega}$ in this region, we calculated ${g_{+-}(r)}$ to a distance of up to ${20}$ times the ionic radius, ${a}$.\\
 \indent Once the values of ${\omega}$ and ${\phi}$ are extracted, we determine the amplitude and correlation length from solution of a least squares problem:
 \begin{equation}
     \arg\min_{(A,\kappa)} \sum_i \left( h_{+-}(r_i) r_i - A\exp(-\kappa r_i) \cos(\omega r_i +\varphi)\right)^{2},
\label{eq:min2}
 \end{equation}
subject to $ \kappa\in(0,\infty) $ and $ A\in[1,\infty) $. Finally, we perform an additional least squares optimization against ${h_{+-}(r)}$ for ${A}$, ${\kappa}$, ${\omega}$, and ${\varphi}$ jointly using simulated annealing and the previously identified best fit values.  Results are summarized in Figure \ref{fig:kappa_omega_gofr}. The extracted ${\kappa}$ and ${\omega}$, normalized by the hard core radius, follow a similar trend as proposed by Adar et al. \cite{Andelman2019} for low ionic strengths. However, consistent with the behavior obtained from the structure factor in Figure \ref{fig:decay_sofq}, the correlation length appears to grow faster than predicted by the shell model with increasing ${\kappa_{\mathrm{D}}}$. \\
\indent There is a spread in the values of ${\kappa}$ and ${\omega}$ when extracted from ${h_{+-}(r)}$ relative to those computed from ${S(k)}$.  However, analyzing the oscillations in real space reveals a jump in the oscillation frequency that is not easily identifiable by analyzing the small ${k}$ behavior of the charge structure factor. As the concentration of ions increases, so does the frequency of oscillations dictated by electrostatic forces. At the points near the dotted orange line in Figure \ref{fig:kappa_omega_gofr} (b), the entropic penalty of the structural configurations is too high to be overcome by electrostatic forces. In these cases, the ion diameter determines the preferred charged oscillation frequency, causing a sudden jump in the values of ${\omega}$ determined from ${h_{+-}(r)}$.  One can see that it is the balance of electrostatic forces and entropic (hard core repulsive) forces that dictates this transition by noting that it occurs at different values of $ \kappa_D a_\mathrm{hs} $ for electrolytes with different strengths of electrostatic interactions $ \epsilon $.  The stronger the electrostatic interactions, the higher the ion concentration required to drive this jump in oscillation frequency.  A charge correlation function depicting oscillations dominated by excluded volume interactions is depicted in Figure \ref{fig:gofr}. This plot shows the mean field model prediction in pink lines, the least square fit to the simulation data in black lines, and the longest wavelength charge oscillations in black markers. Dashed and dotted lines, and filled markers correspond to negative values of ${h_{+-}(r)}$. \\
\indent To further illustrate the competition between steric and electrostatic forces resulting in this jump, let us focus again on ${h_{+-}(r)}$, depicted in Figure \ref{fig:gofr} for a high ionic strength solution and ${a_{\mathrm{hs}}=a}$.  The oscillation period dictated by the mean field prediction (pink line) is approximately ${3a}$.  Over this distance, the charge changes from positive to negative once. If we imagine that this charge oscillation occurs along a line, this suggests an effective inter-ion distance of approximately $ 3 a / 2 $.  For this high ionic strength solution, the volume fraction based on $ a $ is $ 0.45 $.  An effective lineal inter-particle distance can be calculated from the volume fraction of ions in the electrolyte: $ (a^3/\phi_a)^{1/3} \approx 1.3 a = 1.3 a_\mathrm{hs} $, which is smaller than the effective inter-ion distance of $ 1.5a $.  In a sense, the space required to sustain the purely electrostatic charge oscillation wavelength exceeds the available lineal free volume per ion. From this simplified geometric explanation, we see that ion excluded volume forces the ions to rearrange themselves such that they follow the frequency dictated by packing rather than electrostatics. This transition can only be captured if excluded volume interactions are explicitly present in the mathematical formulation of a model of the concentrated electrolyte microstructure. To our knowledge, there are no mean-field theories that capture this structural transition, mainly because charge-charge correlations are represented by a continuous parameter, \cite{BSK2011, Moreira2001} or steric contributions are taken into account by using potential kernels that truncate the range of Coulombic interactions. \cite{Andelman2019}\\
\indent The charge oscillation frequency extracted from $ S( k ) $ in Figure \ref{fig:decay_sofq}, is an approximation that comes from fitting $ S( k ) $ at low values of $ k $ and reporting the pole closest to the real axis. It is assumed that this is the pole that controls the structural behavior of the charge density. This assumption is implicit in equation \ref{eq:hr} by considering a single oscillatory damped exponential. However, equation \ref{eq:hr} is more accurately represented as a summation over relaxation modes:
\begin{equation}
    h_{+-}(r)=\sum_{n=0}^{\infty}\frac{A_{n}\exp(-\kappa_{n}r)}{r}\cos(\omega_{n}r+\varphi_{n}).
    \label{hr_sum}
\end{equation}
The contribution of these other modes is not easily extracted from the small $ k $ fit to $ S(k )$.  Charge oscillations on length scales corresponding to the ion's hard core diameter are reflected in the local maximum (the nearest neighbor peak) in $ S( k )$, which is not incorporated in the identification of the poles.  At low ion packing fractions where there is no nearest neighbor peak, it is safe to assume that the dominant oscillatory mode extracted from $ S(k ) $ at small $ k $ is dictated by electrostatic forces. At high packing fraction, however, there may be structural modes with similar decay lengths but different charge oscillation wavelengths that are difficult to discriminate through asymptotic analysis of $ S( k ) $.  There is value in examining both $ S(k) $ and $ g_{+-}(r ) $.  Long range correlations like the decay are unambiguously observable in $ S( k ) $.  This may also be true of $ g_{+-}( r ) $, but the calculation of this correlation function has computational complexity that scales with the number of ions squared.  Alternatively, $ S( k ) $ computed with non-uniform fast Fourier transformations can have log-linear complexity.  Thus, these calculations are feasible for the large system sizes required to observe correlations over large length scales.
\begin{figure*}[ht]
\includegraphics[width=0.8\textwidth]{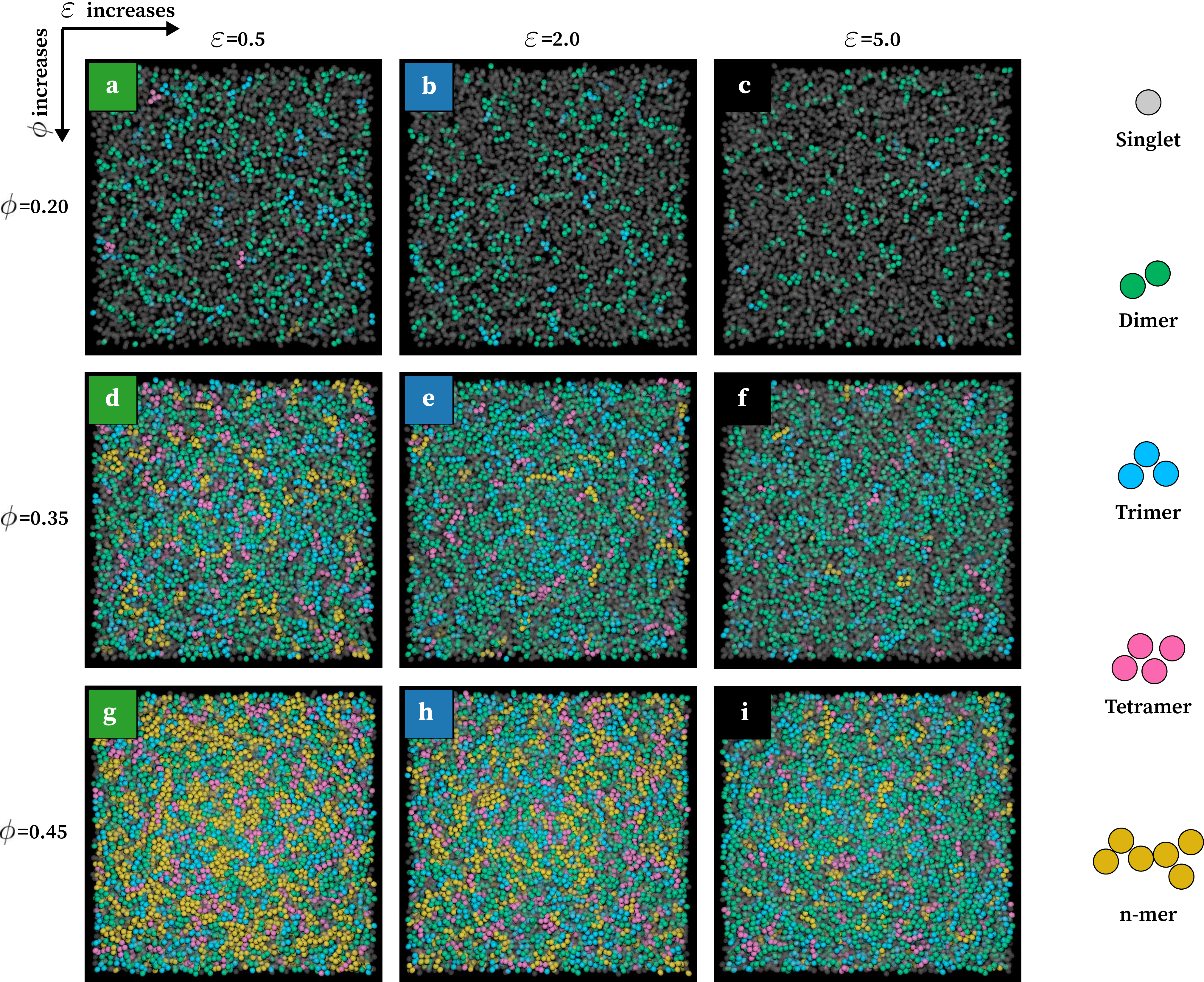}
\caption{\label{fig:snapshots} Snapshots showing the distribution of clusters comprised of same charge ions. Single ions are shown in faded gray, clusters comprised of 2, 3, 4 and 5 or more ions are colored in gray, green, blue, pink and yellow respectively. The strength of the ion-ion interactions increases from left ${\epsilon=0.5}$ to right ${\epsilon=5.0}$. Volume fraction increases from ${\phi=\phi_{a}=0.20}$ to ${\phi=\phi_{a}=0.45}$ downwards. }
\end{figure*}
\subsection{\label{sec:g++_Analysis} Like-Charge Pair Distribution Function and Packing Effects}
\indent The importance of excluded volume forces on electrolyte microstructure also becomes evident if we observe the like-charge pair distribution function ${g_{++}(r)}$. This positive-positive pair distribution function describes the likelihood of finding a positive charge at a distance ${ r }$ from a central positive charge. As our analysis is limited to symmetric electrolytes, this distribution is the same for negative charges (${g_{++}(r)=g_{--}(r)}$). The ${r}$ dependency of the like-charge distribution function is found by direct calculation using equation \ref{eq:binary_gofr}, with $ \nu=\mu $. \\
\indent As depicted in Figure \ref{fig:gqq}, at low concentrations, the ions rearrange themselves such that the probability of finding a similarly charged ion near the central particle is quite low.  $ g_{++}(r) $ is less than 1, which is the value expected of an ideal gas, over a range of $ r / a $ up to as big a value as 5 in some cases.  Like-charges repel and thus the region surrounding an ion is depleted of similarly charged ions relative to an ideal gas. As the concentration of the electrolyte increases and packing effects start to compete with electrostatics, the negative correlation between like-charge ions decreases.  At high enough packing fractions, $ g_{++}(r) $ becomes greater than 1 for $ r / a_\mathrm{hs} \rightarrow 2 $. Even at modest concentrations, packing effects are strong enough for this positive correlation to reach significant values even for weak electrostatic interactions (as in Figure \ref{fig:gqq} (c) where  $\epsilon = 0.5 \: k_{\mathrm{B}}T$). The same occurs with stronger electrostatic interactions but at higher ion concentrations still (as in Figure \ref{fig:gqq} (d) where  $\epsilon = 5.0\: k_{\mathrm{B}}T$). Furthermore, the two rows in Figure \ref{fig:gqq} illustrate how the excluded volume interactions contribute to the like-charge correlations.  The size of peak, in Figure \ref{fig:gqq} (b) and (d), decreases on changing the hard-core radius from ${1.0a}$ to ${0.75a}$. \\
\indent It must be emphasized that the values of $ g_{++}$, shown in Figure \ref{fig:gqq}, are found by averaging over each one of the ions and ${80}$ independent configurations. The fact that like-charges are positively correlated at high concentrations challenges the more intuitive idea that opposite charges will separate as far as possible from each other, and might even tend to arrange themselves in a lattice-like configuration to minimize the potential energy of the system. This simplified view is far from what Figures \ref{fig:gqq} reveals, where, at high concentrations, there is a statistically significant appearance of like-charge structures that could be a result of the presence of complex clusters.%
\begin{figure}
\includegraphics[width=0.48\textwidth]{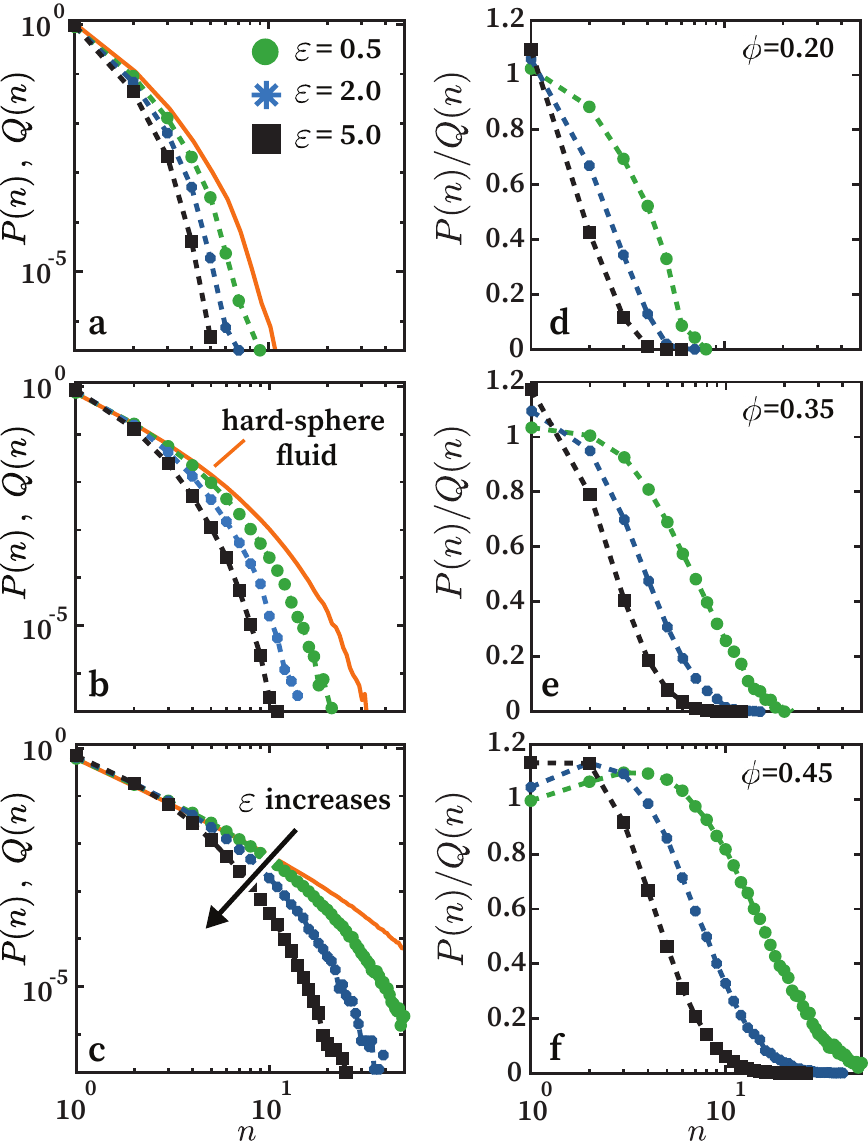}
\caption{\label{fig:distributions} (left) Size probability distribution of like-charge clusters, ${P(n)}$, and the hard-sphere fluid, ${Q(n)}$. Green, blue and black markers correspond to ${\varepsilon=0.5\:k_{\mathrm{B}}T}$, ${\varepsilon=2.0\:k_{\mathrm{B}}T}$, and ${\varepsilon=5.0\:k_{\mathrm{B}}T}$ respectively. Distributions obtained from the hard sphere fluid are shown by the continuous orange line. Volume fraction increases downwards from ${\phi=0.20}$ (first row) to ${\phi=0.45}$ (third row). (right) Figures in the second column show the evolution of the distribution of cluster sizes, ${P(n)}$, normalized by the distribution from a binary hard sphere fluid at the same volume fractions, as ${\epsilon}$ and ${\phi}$ vary. Each of the rows correspond to the rows of Figure \ref{fig:snapshots}.}
\end{figure}
\subsection{\label{sec:Cluster_Analysis} Like-charge Cluster Analysis}
\indent As indicated in the previous sections, analysis of the electrolyte pair's structural properties at various volume fractions, ionic strengths, and hard-sphere radii suggests the formation of complex clusters comprised of equally charged species. To investigate this phenomenon, we performed a cluster analysis of like-charges in solution. In this study, two ions are said to belong to the same cluster if they share the same charge, and the distance between the ions is less than ${ 2.1 a_{\mathrm{hs}} }$. We chose this distance to only include the ions that reside within the first shell of neighbors. We extracted the probability distribution of the cluster size, ${P(n)}$, where ${n}$ is the number of like-charges that belong to the same cluster, and sampled from $ 100 $ configurations equally distanced at time intervals for which density fluctuations are decorrelated.\\
\indent There are two notable weaknesses in this approach. First, the introduction of a user-defined geometrical length or cut-off radius might bias the sample. Second, there is no simple mathematical treatment that allows us to simply identify the driving forces that form these clusters. A practical way to circumvent these limitations is to introduce a ``reference state'' or, more precisely, a reference probability distribution. We will denote this distribution as ${Q(n)}$, and define it to be the probability distribution of cluster sizes obtained from simulations of the exact same electrolyte geometrically but without any electrostatic interactions among the ions, as though the temperature tends to infinity. We posit that by analyzing the clusters in this way, we can reduce any bias from the cut-off radius, and distinguish the role of electrostatic interactions from packing effects in clustering of like-charges. \\
\indent The size distribution of these clusters is shown in Figure \ref{fig:snapshots}, represented through sample configurations and quantitatively in Figure \ref{fig:distributions}. The first three columns show snapshots at different strengths of electrostatic interaction: ${\varepsilon=[0.5,2.0,5.0]}$, while each of the rows corresponds to different volume fractions:  ${\phi=\phi_{a}=[0.20,0.30,0.45]}$. Like-charge clusters are colored by size. Ions that do not belong to a cluster are gray and muted out to facilitate the visualization of the clusters.\\
\indent The importance of ion concentration is clear. At relatively low concentrations (first row), clusters are mainly comprised of two or three ions, and most are a single ion. As the concentration increases, larger like-charge clusters form, and a larger fraction of ions belongs to a cluster. The strength of electrostatic interactions has the opposite effect.  With increasing $ \epsilon $, the number of ions forming larger clusters decreases. This behavior can be explained by the competition of two opposing forces. Electrostatic forces disfavor like-charge clustering, but excluded-volume forces do not. \\
\indent The snapshots in Figure \ref{fig:snapshots} show qualitatively the effect that varying the concentration and ${\varepsilon}$ have on the proliferation of like-charge clusters. This dependency is also quantified as in Figure \ref{fig:distributions}. The first column depicts the probability distribution of a cluster comprised of ${ n }$ like-charge ions, ${P(n)}$. The second column is the same probability normalized with respect to the cluster size distribution obtained from the same electrolyte but with $ \epsilon = 0 $, denoted ${Q(n)}$. This reference configuration is essentially a hard sphere fluid with labeled particles. Values of ${P(n)/Q(n)>1}$ indicate that clusters of size ${n}$ in the charged system are more prevalent than in the uncharged one, and the contrary is true for values of ${P(n)/Q(n)<1}$. Each of the plots of probability and probability ratio shares its ion concentration in common with the respective row of the plot. That is, for ${a_{\mathrm{hs}}=1.0a}$, ${\phi=\phi_a = [0.20,0.35,0.45]}$ for the first, second and third rows respectively.
\indent At concentrations of ${\phi=0.20}$ (first row), the observed behavior is what one might expect. There is a higher probability of like-charge ions being isolated in the electrolytes relative to the hard sphere fluid. Increasing the strength of the ion-ion interactions at fixed ion concentration only accentuates this trend. However, the ratios calculated at higher ion concentrations reveal something different. The plotted ratio in the middle row ($ \phi = 0.35$) shows that dimers for the electrolyte with the lowest value of ${\varepsilon}$ (green markers) are more prevalent than in the hard sphere fluid. The ratio plotted in the last row, where ${\phi =0.45}$, shows a higher probability of observing clusters comprised of a small number of ions: 2-8, than what is expected of a hard-sphere liquid for the smallest value of $ \epsilon $. Notably, for this lowest strength of electrostatic interactions, clusters comprised of a small number of ions (2-6 elements) have not only a higher probability of being present in the charged system, but their odds ratio, ${P(n)/Q(n)}$, is also higher than for finding singlets. There is an overabundance of like-charge clusters. While this behavior is less prominent as $\varepsilon$ increases further, it is still evident at the highest ion concentration depicted. When ${\varepsilon=2.0}$ (blue markers), there is an overabundance of clusters containing two and three like-charges. When $ {\varepsilon = 5.0} $ (black markers), the odds ratios for singlets and doublets are essentially equal and still greater than unity.  For larger clusters, this odds ratio falls well below that mark.\\
\begin{figure}[t]
\includegraphics[width=0.45\textwidth]{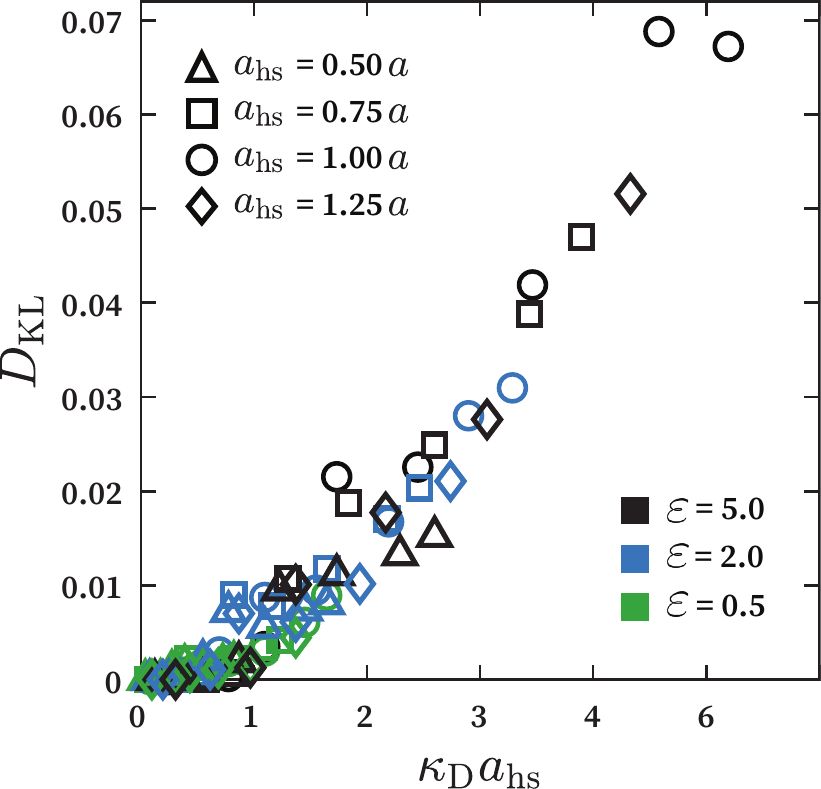}
\caption{ \label{fig:KL_div} 
Kullback-Leibler divergence of the cluster size distribution for a symmetric binary electrolyte ${P(n)}$, with respect to the cluster size distribution for a binary hard sphere mixture at the same volume fraction, ${Q(n)}$, as a function of the inverse of the Debye length. }
\end{figure}
\indent While some clear trends can be obtained from the probability distributions, as shown in figure \ref{fig:distributions}, mapping  the difference between the distributions into a scalar quantity or a ``statistical distance'' could facilitate the interpretation of these trends. Information theory gives us a method to quantify the disparity of a distribution ${P(n)}$ from some reference distribution ${Q(n)}$ through the Kullback-Leibler (KL) divergence, also known as relative entropy. Here we use these two terms interchangeably. The KL divergence is defined as:\cite{kullback_1997}
\begin{equation}
    D_{ \mathrm{KL} } (P\parallel Q)= \sum_{x\in { \mathcal{X} } }P(x)\log \left({\frac{P(x)}{Q(x)}}\right)\geq0 \; ,
    \label{eq:KL}
\end{equation}
where ${\log(x)}$ denotes the natural logarithm of ${x}$. Non-negativity of this quantity follows from Jensen’s inequality\cite{Altaner2017}, and, in the simplest interpretation, it is only zero if ${P(x)=Q(x)}$.\\
\indent Notice that when ${Q(x)}$ is a uniform distribution, equation \ref{eq:KL} differs from the Shannon entropy of the distribution $ P( x ) $ by a constant. This analogy to entropy can be useful for interpreting the KL divergence.  For the distribution of like-charge clusters, $ Q( n ) $, we can imagine a symmetric electrolyte at a given concentration and with zero charge.  The distributions $ P( n ) $ in the different left panels of Fig. \ref{fig:distributions} are those found when increasing the magnitude of the ion charges until they reach a value ${\pm q}$. We should expect that the measure of entropy associated with the like-charge cluster size distribution will change because the distribution of electric potential throughout the electrolyte changes.\\
\indent If probabilities ${P(x)}$ and ${Q(x)}$ are Boltzmann distributed, the KL divergence has a thermodynamical meaning. Suppose we have a system (with constant volume) in equilibrium with a reservoir at ${T=T_{1}}$, and we put it in contact with a bath at ${T=T_{2}}$. The system's entropy will change by ${\Delta S=\left\langle S\right\rangle _{2}-\left\langle S\right\rangle _{1}}$, where subscripts $1$ and $2$ in$\left\langle \cdots\right\rangle$ indicate the ensemble averages at $T_1$ and  $T_2$, respectively. The entropy of the final bath will change by $\Delta S^{\text{bath}}=(\left\langle U\right\rangle _{2}-\left\langle U\right\rangle _{1})/ T_{2}$ (the heat exchanged between the system and the heat bath). The sum of these contributions equals the KL divergence. Therefore, the thermodynamic interpretation of the relative entropy applied to both distributions represents the total change in entropy due to putting our system in contact with the bath. Additionally, if the energy exchange between the system and the second bath is performed via a Carnot engine, the KL divergence would provide an upper bound to the maximum work that could be extracted from the energy exchange in units of entropy.\cite{Gaveau2014}\\
\indent The calculated KL divergence as a function of ${\kappa_{\mathrm{D}}} a_\mathrm{hs}$ is shown in Figure \ref{fig:KL_div} for all the electrostatic strengths, hard core sizes, and ion concentrations employed in this study. When ${\kappa_{\mathrm{D}}}$ is scaled on the electric radius, ${a}$, we found no evident trend. However, when plotted this way ($\kappa_{\mathrm{D}}$ scaled on $a_\mathrm{hs}$), the KL divergence lies nearly on a universal curve.  We can observe that for large Debye lengths (low ionic strengths), the KL divergence tends towards zero (on the order of ${10^{-4}}$). This indicates that deviations from the binary-hard-sphere-liquid cluster size distribution are minimal.  We attribute this to electrostatic screening combined with a large free volume available to each ion. Larger differences between ${P(n)}$ and ${Q(n)}$ are present as ${\lambda_{\mathrm{D}}^{-1}}$ increases. For these shorter Debye lengths, a competition between electrostatic interactions and excluded volume interactions leads to an overabundance of small like-charge clusters with respect to the labeled hard sphere liquid. Most noticeably, for values of ${\kappa_{\mathrm{D}}a_{\mathrm{hs}}>1}$ (also where the onset of oscillations appears in figure \ref{fig:decay_sofq}), the KL divergence appears to grow monotonically with the inverse of the Debye length. Nonetheless, the presence of statistical noise for low values of ${\kappa_{\mathrm{D}}a_{\mathrm{hs}}}$ presents difficulties in precisely determining the scaling of this monotonic increase in the relative entropy as ${\kappa_{\mathrm{D}}a_{\mathrm{hs}}}$  increases.\\
 \indent As ${\kappa_{\mathrm{D}}}$ is inversely proportional to the temperature, Figure \ref{fig:KL_div} can be interpreted as the relative entropy of the two probability distributions as the temperature decreases. At high temperatures, electrostatic forces, in the case of the charged mixture, can be considered to be negligible with respect to the kinetic energy. The kinetic energy of both the charged and the binary hard-sphere liquid determines the cluster size probability distributions. Thus, there is little difference between the distributions. Nonetheless, as the temperature decreases (${\kappa_{\mathrm{D}}}$ increases), electrostatic forces play an important role in determining the cluster size probability distributions, thus increasing the distance between the two probability distributions. \\
\indent Qualitatively, the ratio of the distributions shown in Figure \ref{fig:distributions} suggest that ${ P(n) }$ differs more from ${ Q(n) }$ as the strength of electrostatic interactions, $ \epsilon $, increases (or equivalently, ${ \kappa_{ \mathrm{D} } }$ at fixed ion concentration decreases). In contrast, the computed KL divergence shows that the distance between the two distributions increases with the inverse of the Debye length. The reason for this apparent contradiction is explained by a difference in the weights used in computing the KL divergence. The ratios of distributions in Figure \ref{fig:distributions} are unweighted, whereas the KL divergence shown in Figure \ref{fig:KL_div} uses ${ P(n) }$ as the weight of the logarithmic difference, resulting in contributions to the KL divergence that are inversely proportional to the cluster size. It is the smaller like-charge clusters that contribute most to the KL divergence. \\
\indent Overall, in this analysis, we have gained valuable information about the importance of electrostatic interactions on the existence of these like-charge clusters. Figure \ref{fig:distributions} shows that their size is highly correlated to the strength of electrostatic interactions, ${\varepsilon}$. Furthermore, it can be inferred that there is an excess of small clusters with respect to the hard-sphere fluid at high concentrations and that they contribute the most to the statistical distance (KL divergence) between the distributions ${P(x)}$ and ${Q(x)}$. The KL divergence (Figure \ref{fig:KL_div}) shows that, even for many-body correlations, properties appear to fall nearly on a universal curve when ${\kappa_{\mathrm{D}}}$ is scaled on the hard-sphere radius, and numerically significant differences between the cluster size distributions are present for values of ${\kappa_{\mathrm{D}}a_{\mathrm{hs}}>1}$.
\section{\label{sec:Conclusions} Conclusions}
\indent In this work, we use Brownian Dynamics coupled with a truncated multipole expansion of the electric potential to study the structure of concentrated electrolytes using a version of the Restrictive Primitive Model. We focus on the decay of spatial correlation functions and extract the inverse of the correlation length, ${\kappa}$, and frequency of oscillation, ${\omega}$. By varying the radius of the hard-core repulsion, we investigate the importance of packing effects at high ion concentrations. The analysis is performed over a large range of ion concentrations, extending from the very dilute regime, where the Debye-H\"uckel theory is applicable, to the regime where ionic hard-sphere packing dominates. We use a simple scheme to extract ${\kappa}$ and ${\omega}$ from the charge structure factor ${S(k)}$ at small ${k}$, and compare their values against those obtained by sampling the charge-charge correlation function, ${h_{+-}(r)}$, using the standard ``shell summation'' method. Figure \ref{fig:decay_sofq} (b) and \ref{fig:kappa_omega_gofr} (c) show the computed correlation lengths as a function of the inverse of the Debye length, ${\kappa_{\mathrm{D}}}$. The obtained scalings of ${(\kappa_{\mathrm{D}}/\kappa)\sim \kappa_{\mathrm{D}}^{n}}$, with ${n=2.0}$ and ${n=1.8}$ are higher than those found by recent computational studies \cite{coles_rotenberg2020,Andelman2019,Cats_VanRoij2020} in the range of physically feasible values of ${\kappa_{\mathrm{D}}}$. Similar values of n have only been found in computational studies by fitting the radial distribution to the sum of two relaxation modes and extracting two distinct correlation lengths that scale as $n_1 \approx 1$ and $n_2 \approx 2$, respectively.\cite{zeman_holm2020} We do not, however, obtain the experimentally observed power law scaling for which ${n=3}$. \cite{Gebbie2017} Perhaps, there are other important many-body physics leading to alterations of the ion microstructure in the neighborhood of macroscopic boundaries that produces more underscreening than observed in the charge correlations for bulk electrolytes.  We also find a sudden jump in the values of ${\omega}$ obtained from ${h_{+-}(r)}$ that is not well captured in the asymptotic analysis of ${S(k)}$. We attribute this structural transition to excluded volume interactions that are, by definition, not captured in the far-field decay of the correlation functions, and commonly not taken into account in mean-field models.  This jump does not appear to affect the power law decay of the charge correlations in any significant way.\\
\indent Furthermore, we study the correlations between like-charges. Figure \ref{fig:gqq} shows that like-charges transition from being negatively correlated to positively correlated as the ion concentration increases, suggesting the existence of complex clusters of charge rather than single or paired ions. We believe this emergent behavior to be entropically driven, and use the simulated configurations to identify clusters of ions with the same signed charge. Figure \ref{fig:snapshots} shows qualitatively and Figure \ref{fig:distributions} quantitatively the size distribution of these clusters. The cluster size distributions in the second column of Figure \ref{fig:distributions} are normalized with respect to simulations of the same system without electrostatic interactions among the ions. We use this ``reference state'' to reduce any bias introduced by the cut-off radius used to identify clustered ions and to distinguish the role of electrostatic interactions from packing effects. We validate the positive correlations in $ g_{++}(r) $ at short ranges corresponding to like-charge ion clusters, and find that clusters composed of a small number of ions (less than 10 ions) are more prevalent in the electrolyte than in the simple binary hard-sphere simulations. The analysis reveals clear trends in the cluster size distribution.  Packing promotes like-charge clusters, while electrostatic interactions inhibit their formation.\\
\indent We use the Kubler-Leibler divergence to compute the relative entropy of the cluster size distributions with and without electrostatics.  We find that it is near zero at low values of ${\lambda_{\mathrm{D}}^{-1}}$, indicating that there is little entropic difference between the two systems. Nonetheless, once the screening length is in the order of the ion diameter, the relative entropy increases monotonically. There are two reasons for this deviation in relative entropy from zero. \\
\indent In this work, we have have seen that the size of the hard-core represents an important physical length scale in determining the structural properties of a concentrated electrolyte. Evidence of this can be observed not only by the computed correlation lengths and frequencies of oscillation, but also the relative entropy for like-charge clusters.  These measures of the microstructure appear to reliably collapse on master-curves when ${\kappa_\mathrm{D}}$ is scaled on ${a_{\mathrm{hs}}}$.  Understanding why this collapse occurs is important.\\
\indent Consider the case of two interacting ions in solution. The strongest Coulomb interaction, $ U_\mathrm{max} $, that the ions experience when $ a_\mathrm{hs} > a $ scales as $ q^2 / a_\mathrm{hs} $.  When this energy scale is normalized by the thermal energy, $ k_{\mathrm{B}} T $, it can be rewritten as $ U_\mathrm{max} / k_{\mathrm{B}}T \sim z^2 \lambda_\mathrm{B} / a_\mathrm{hs} $, the  characteristic energy scale for Coulombic interactions. This is not of particular concern in the case of point charges because they occupy an infinite small volume that allows the charges in the lattice to have a minimum distance of closest approach. Nonetheless, if hard-core repulsions, which are present in reality, are taken into account from the ``swelling'' process, the interparticle distance dictated by the effective charge is shorter than the diameter of the swelled ions. This renders the interparticle distance that would minimize the internal free energy for point charges no longer an accessible state. Thus, the length scale that controls the structural properties of concentrated electrolytes is not the physical length scale that naturally arises from electrostatic interactions, but rather the physical length scale that arises precisely from the non-continuum nature of the hard-core repulsions, and commonly not considered in mean-field theories.\\
\indent Additionally, this work exposes some hidden structures that emerge in concentrated electrolytes. It seems that collective effects might be important in understanding and predicting underscreening. The presence of like-charges in the neighborhood of a central ion, differs the commonly understood picture of screening, as there are not sufficient counterions to ``shield'' the emanated electric field.  Counterintuitively, there are charges with equal sign that locally enhance the gradient of the electric potential. This could have additional important theoretical consequences, because it suggests that concentrated electrolytes may have more heterogeneity in the local charge distribution and exhibit higher local charge densities than anticipated by naive mean field theories.  There is a distribution of like-charge clusters that affects the local electrolyte structure and might influence the effective screening of electrostatic forces between macroscopic charged surfaces in ways that differ from that observed for the charge pair correlation function.\\ 
\indent We leave it to future work to quantify how fluctuations in local charge density due to the existence of like-charge clusters and additional topological constraints from the charged surfaces in surface force apparatuses affect bulk screening. We think this might be able to explain the discrepancy between the observed scalings for the screening length found by experiments using macroscopic surfaces as transducers of electrostatic forces and those predicted by computational methods for bulk electrolytes. While experiments to study these clusters in concentrated electrolytes are technically difficult because of the length scales involved, similar conditions could be obtained in a suspension with oppositely charged colloids. These kind of colloidal systems have been previously used to test the phase behaviour predicted by the Restrictive Primitive Model. \cite{Hynninen2006} The local structure in such a colloidal electrolyte could be observed with fluorescence microscopy.\cite{Leunissen2005,ElMasri2012} 
\begin{acknowledgments}
We wish to acknowledge the support from NASA, Grant No. 80NSSC18K0162 and NSF, Career Award No. 1554398.
\end{acknowledgments}
\section*{Author Declarations}
\subsection*{Conflict of interest}
The authors have no conflicts to disclose.
\section*{Data Availability Statement}
The data that support the findings of this study are available from the corresponding author upon reasonable request.
\appendix
\section{\label{sec:Relaxation} Equilibration process and structural relaxation }
\begin{figure}[ht]
\includegraphics[width=0.48\textwidth]{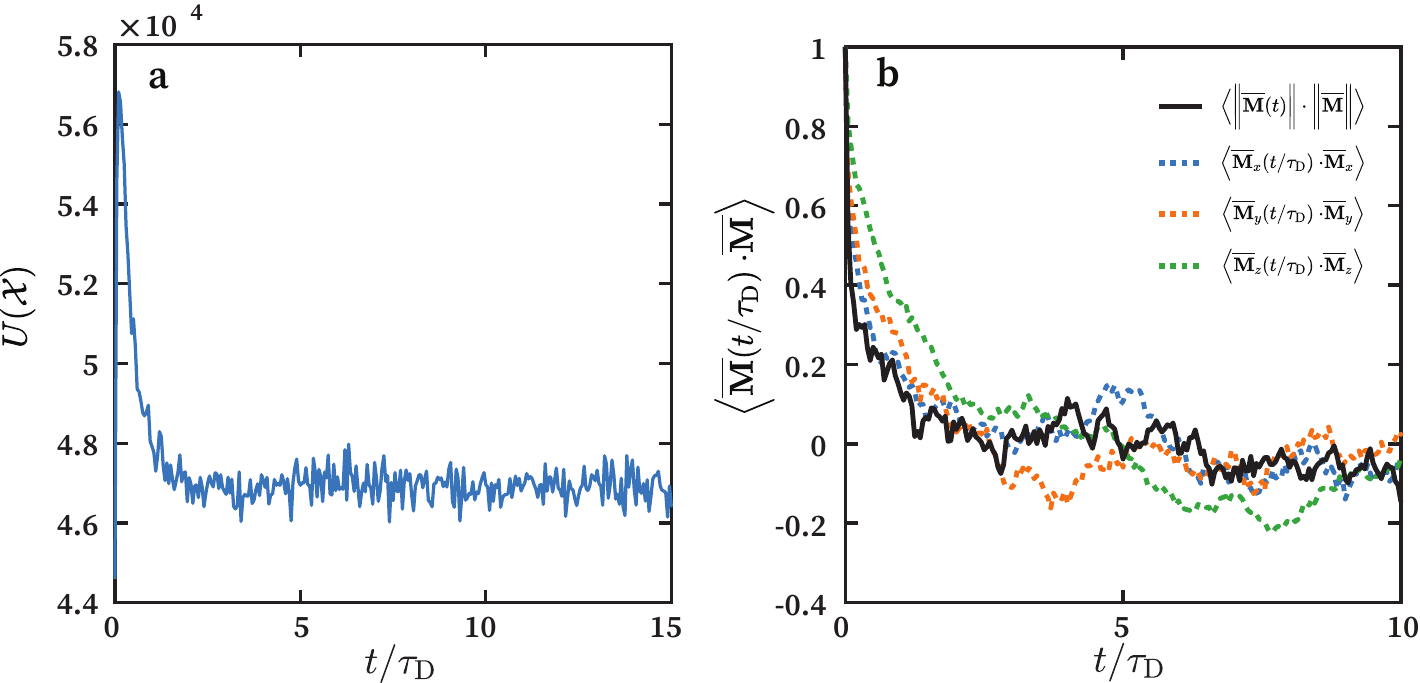}
\caption{ \label{fig:relaxation} ( \textbf{a} ) Time evolution of the total potential energy and ( \textbf{b} ) the dipole auto-correlation function for the case of ${\varepsilon=5.0}$, ${\phi=\phi_{a}=0.45}$. The auto-correlation function is shown for the magnitude of the dipole moment (black continuous line) and for each of its Cartesian components; the ${x}$, ${y}$ and ${z}$ components are presented by the blue, orange and green dotted lines respectively. }
\end{figure}
\indent The equilibration process is as follows: first, we randomly place all labelled (positive and negative) ions in the simulation box. The box length is of equal magnitude in all simulations (${L=100a}$). Then, simulations are run for ${50\tau_{\mathrm{D}}}$ without any electrostatic interactions. Finally, electrostatic interactions are enabled and the simulations are run for an additional ${150\tau_{\mathrm{D}}}$. The first ${100\tau_{\mathrm{D}}}$ (${50\tau_{\mathrm{D}}}$ without electrostatic interactions and ${50\tau_{\mathrm{D}}}$ with electrostatic interactions) are discarded.\\
\indent We tracked the total potential energy and the total dipole moment of the simulation box to validate the previous equilibration steps as sufficient. We computed the normalized dipole moment auto-correlation function:\cite{Macquarrie2000}\\
\begin{equation}
\left\langle \overline{\mathbf{M}}(t)\cdot\overline{\mathbf{M}}(0)\right\rangle =\frac{\left\langle \mathbf{M}(t)\cdot\mathbf{M}(0)\right\rangle }{\left\langle \mathbf{M}(0)\cdot\mathbf{M}(0)\right\rangle }\; ,
\end{equation}
where $\mathbf{M}(t)$ is the total dipole moment of the simulation box:\\
\begin{equation}
    \mathbf{M}(t)=\sum_{\alpha}q_{\alpha}\cdot \mathbf{x}_{\alpha}(t)\;
\end{equation}
where ${\alpha}$ runs over all particles. This particular auto-correlation function is associated with the structural (or dielectric) relaxation of the entire simulation box.\cite{Macquarrie2000}\\
\indent The evolution of the total potential energy after the electrostatic interactions are enabled and the normalized dipole moment auto-correlation function are shown in Figure \ref{fig:relaxation} (a) and (b) respectively. Both figures correspond to the case of ${\epsilon=5.0}$ and ${\phi_{a}=\phi=0.45}$. Subfigure (a) shows that the potential energy fluctuates around a stable value within ${5\tau_{\mathrm{D}}}$. Similarly, the dipole moment auto-correlation function (subfigure (b)) decays to zero within ${5\tau_{\mathrm{D}}}$. Given that the discarded time is one order of magnitude greater than the relaxation times of the simulation box, we can conclude that the averaged quantities in this research correspond to equilibrium properties. 
\section{Box size dependence of the like-charge cluster probability distribution.}
\begin{figure}[t]
\includegraphics[width=0.45\textwidth]{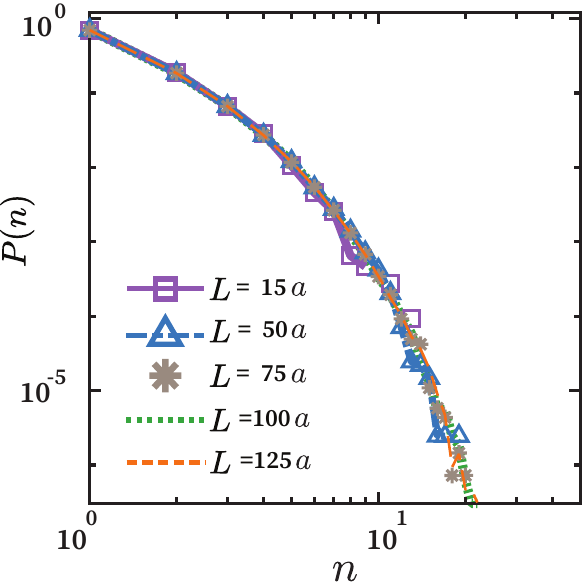}
\caption{ \label{fig:dPndV} Cluster size probability distribution ${P(n)}$ for the case of ${\epsilon=5.0}$ and ${\phi_{a}=\phi=0.45}$ found for different box lengths ${L=[15a,50a,75a,100a,125a]}$. Notice that the cluster size probability distributions no system size dependence for the cases of ${L\geq75a}$. 
}
\end{figure}

\indent The cluster size distribution ${P(n)}$ was calculated for different box sizes to ensure that the like charge clusters are not an artifice of the periodic boundary conditions in the simulation. Figure \ref{fig:dPndV} shows the case of ${\epsilon=5.0}$ and ${\phi_{a}=\phi=0.45}$ found for different box lengths ${L=[15a,50a,75a,100a,125a]}$. The plot shows little to no difference in the behavior of ${P(n)}$ that correspond to box lengths of ${L=75a,100a,125a]}$ (gray asterisks, dotted green line and dashed orange line respectively). In contrast, box sizes corresponding to $L=15a$ and $L=50a$ show slight deviations from the other distributions (${L\geq75a}$) for ${n>8}$, which probably result from insufficient numbers of ions in the simulation to enable accurate sampling of large clusters.\\
\indent As no differences in the distribution are present for ${L\geq75a}$, we conclude that the box length size equal to ${100a}$ used for all simulations discussed in our manuscript is sufficient to prevent any artifice the periodic boundary conditions might otherwise introduce. 

%

%

\end{document}